\documentclass[10pt]{article}

\usepackage{fullpage}
\usepackage{setspace}
\usepackage{parskip}
\usepackage{titlesec}
\usepackage[section]{placeins}
\usepackage{xcolor}
\usepackage{breakcites}
\usepackage{lineno}
\usepackage{hyphenat}

\PassOptionsToPackage{hyphens}{url}
\usepackage[colorlinks = true,
            linkcolor = blue,
            urlcolor  = blue,
            citecolor = blue,
            anchorcolor = blue]{hyperref}
\usepackage{etoolbox}

\usepackage{natbib}

\renewenvironment{abstract}
  {{\bfseries\noindent{\abstractname}\par\nobreak}\footnotesize}
  {\bigskip}

\titlespacing{\section}{0pt}{*3}{*1}
\titlespacing{\subsection}{0pt}{*2}{*0.5}
\titlespacing{\subsubsection}{0pt}{*1.5}{0pt}

\usepackage{authblk}

\usepackage{graphicx}
\usepackage[space]{grffile}
\usepackage{latexsym}
\usepackage{textcomp}
\usepackage{longtable}
\usepackage{tabulary}
\usepackage{booktabs,array,multirow}
\usepackage{amsfonts,amsmath,amssymb}
\providecommand\citet{\cite}
\providecommand\citep{\cite}

\newif\iflatexml\latexmlfalse
\providecommand{\tightlist}{\setlength{\itemsep}{0pt}\setlength{\parskip}{0pt}}%

\AtBeginDocument{\DeclareGraphicsExtensions{.pdf,.PDF,.eps,.EPS,.pdf,.pdf,.tif,.TIF,.jpg,.JPG,.jpeg,.JPEG}}
\makeatletter

\usepackage[utf8]{inputenc}
\usepackage[english]{babel}

\usepackage{float}

\usepackage[margin=1.5in]{geometry}

\begin{document}

\title{The mathematics of contagious diseases and their limitations in forecasting}

\author[1]{Carlos Oscar S. Sorzano}%
\affil[1]{Natl. Center of Biotechnology (CSIC), Madrid, Spain}%

\vspace{-1em}

  \date{}

\begingroup
\let\center\flushleft
\let\endcenter\endflushleft
\maketitle
\endgroup

\selectlanguage{english}
\begin{abstract}
This article explores mathematical models for understanding the evolution of contagious diseases. The most widely known set of models are the compartmental ones, which are based on a set of differential equations. But these are not the only models. This review visits many different families of models: deterministic compartmental models based on differential equations, stochastic processes, stochastic differential equations, Gaussian process regression, Bayesian regression, agent-based models, and integro-differential models. In this regard, this is one of the most comprehensive reviews. Additionally, we show these families, not as unrelated entities, but following a common thread in which the problems or assumptions of a model are solved or generalized by another model. In this way, we can understand their relationships, assumptions, simplifications, and, ultimately, limitations.

Prompted by the current Covid19 pandemic, we have a special focus on spread forecasting. We illustrate the difficulties encountered to do realistic predictions. In general, they are only approximations to a reality whose biological and societal complexity is much larger. Particularly troublesome are the large underlying variability, the problem's time-varying nature, and the difficulty to estimate the required parameters for a faithful model. This complexity makes that predictions based on simple models are not very useful in the long term. Additionally, we will also see that these models have a multiplicative nature implying that small errors in the system parameters cause a huge uncertainty in the prediction. Stochastic or agent-based models can overcome some of the modeling problems of systems based on differential or stochastic equations. The main difficulty in using these models is that they are as accurate and realistic as the data available to estimate their detailed parametrization, and very often this detailed data is not at the modeller's disposal. 

Although the predictive power of mathematical models to forecast the evolution of a contagious disease is very limited, these models are still very useful to plan interventions as they can calculate their impact if all other parameters stay fixed. They are also very useful to understand the properties of disease propagation in complex systems.

\end{abstract}%

\sloppy

\section*{Article Type}

\begin{itemize}
\tightlist
\item
  Review article
\end{itemize}

\section*{Authors}

{\label{290010}}

Carlos Oscar S. Sorzano $^*$\\
ORCID: 0000-0002-9473-283X\\
Natl. Center of Biotechnology, CSIC\\
28049 Madrid, Spain\\
coss@cnb.csic.es\\

The author declares not to have any conflict of interest.

\section*{Graphical/Visual Abstract and
Caption}

\begin{center}
    \includegraphics[width=0.65\textwidth]{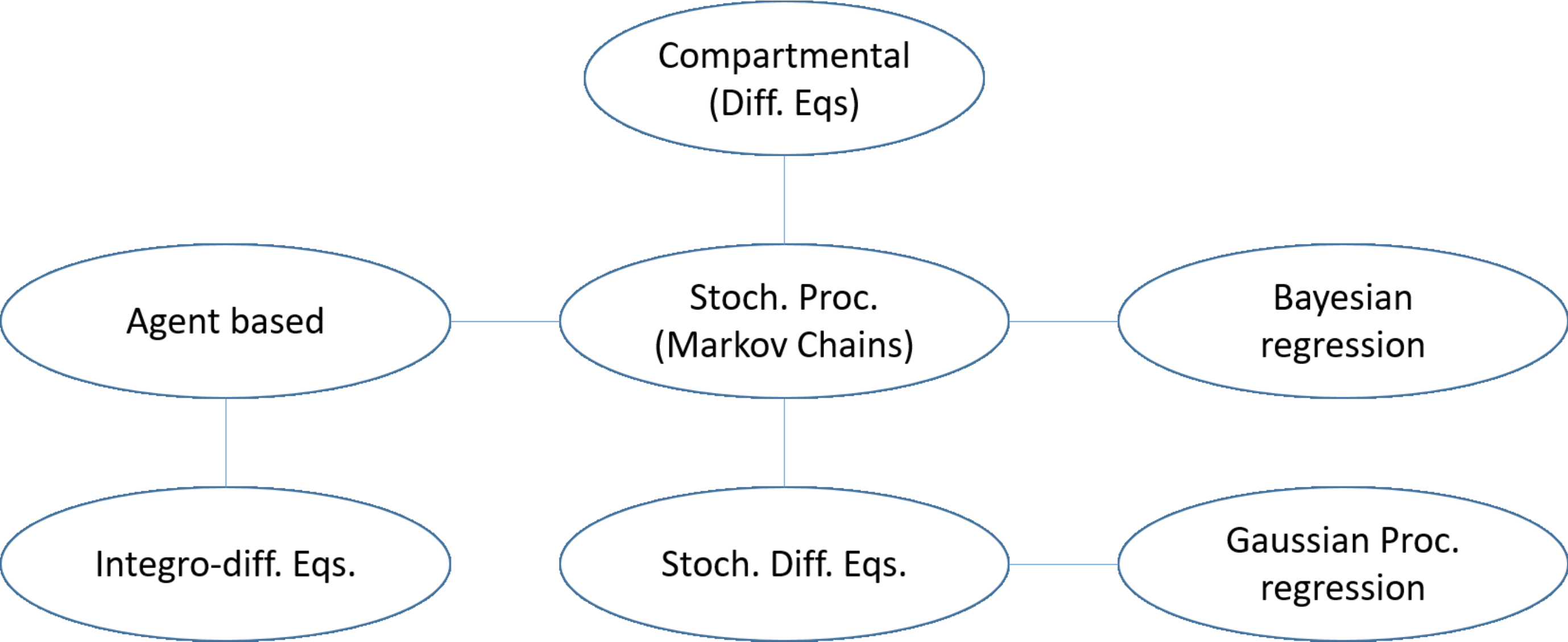}\\
\end{center}

Relationships between the different families of disease spread models.

\par\null

\section{Introduction}

\begin{figure}
    \centering
    \includegraphics[width=8cm]{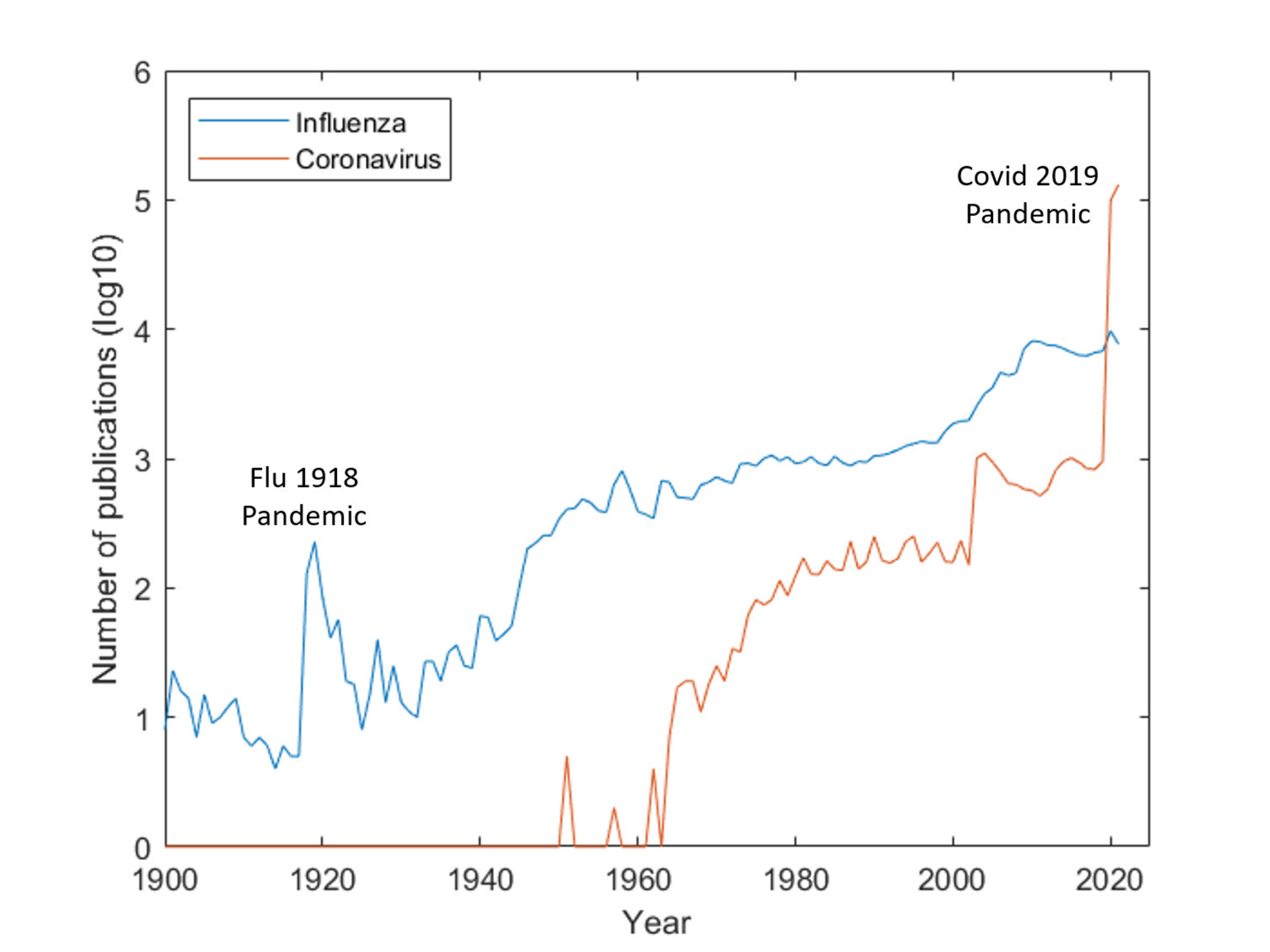}
    \caption{Number of scientific publications about influenza and coronavirus (in log10 units) according to Scopus since the beginning of the XXth century. We have marked the pandemics of influenza in 1918 and Covid19 in 2019. Note that in 2021 there has been an order of magnitude more papers about Covid19 than influenza, that kills worldwide about 0.5M people every year. Also, the number of publications on coronavirus has increased by two orders of magnitude in a single year.}
    \label{fig:influenza}
\end{figure}

2020 has unfortunately been known for a world pandemic with huge human, social, and economic consequences. Science has, more than ever, been brought into focus as the primary source of solutions to isolate the pathogen, track its spread and its evolution, design measures to avoid its propagation, find a cure and a vaccine, etc. Covid19 has been probably the most studied pandemic in history (see Fig. \ref{fig:influenza}), and this study has occurred, and is still happening, in real-time as new data is available. Research has not been restricted to the biomedical aspects of the pathogen and the disease, but it has also covered any societal and economical aspect. In respect to the disease spread, news media have brought to the general public concepts like the basic reproduction number of epidemiological models and, in general, society has become aware of the importance of mathematical modeling as a way to understand the evolution of the pandemics. 

Maybe compartmental epidemiological models of contagious diseases are the most widely known models. They are close relatives of the standard system analysis approach based on the description of the evolution of a deterministic system using differential equations. The system state would be described by internal variables that evolve as prescribed by a differential equation system. The simplest of these models would consider three internal variables: 1) the fraction of the population at time $t$ that is susceptible to be infected, $s(t)$; 2) the fraction that can infect other individuals, $i(t)$; and 3) the fraction that is removed from the dynamics either because they have recovered from the disease and have developed immunity or because they have died from it and cannot propagate it further, $r(t)$. This is the well-known SIR model, and it can be used for quick outbreaks like the one of the Covid19 \citep{Cooper2020}. Once we know the initial state of the system, its evolution is determined by the following differential equations:
\begin{equation}
    \begin{array}{rcl}
         \frac{ds}{dt}&=&-\beta is \\
         \frac{di}{dt}&=&\beta is - \gamma i \\
         \frac{dr}{dt}&=&\gamma i \\
    \end{array}
    \label{eq:SIR}
\end{equation}
with the constraints that the population size remains constant and $s+i+r=1$. Two terms govern these equations: $\beta is$ and $\gamma i$. If $\gamma$ is the recovery rate, then the latter term represents the fractional rate of recovery of infected people. These individuals disappear from $i$ and appear as $r$. Similarly, given that at a particular time we have $s(t)$ susceptible people and $i(t)$ infected people, assuming that these people are homogeneously mixed, the probability of an encounter is proportional to the product $is$ (if there are very few susceptible or infected people, this probability is very small). $\beta$ is the infection rate, that is, a number that encompasses all the effects involved in the infection process (probability of encounter, probability of an encounter resulting in an effective infection, ...). At this point, it is clear the effect of measures like the ``social distance'' or lockdown. The goal is to reduce this infection rate as much as possible. It should also be noted that $\beta$ and $\gamma$ are not purely biological characteristics of the pathogen and its host. They are related to the spread of the disease and its recovery or death toll. Consequently, they also depend on the different countries' economic and health systems and their environmental pollution. Still, we advance here that the transmission and recovery processes are rather complicated events and cannot be reduced to a single rate. In an extremely simplified model, we could think of $\beta$ as the product between the average number of contacts of an infectious individual and the probability of a contact being infected. In this simple model, $\gamma$ would be the inverse of the average time to stop being infectious (either because the person has recovered from the disease or because he/she has died from it).

The basic reproduction number is defined for this simple model as
\begin{equation}
    R_0=\frac{\beta}{\gamma}
    \label{eq:R0}
\end{equation}
\citet{Katul2020} and \citet{Odriscoll2021} shows how to estimate the basic reproduction number in the early stages of an epidemic. If $R_0<1$, then the infectious pool, $i$, depletes more quickly than it is filled with new infections, and the epidemic disappears soon due to the lack of encounters between infected and susceptible people. On the contrary, if $R_0>1$, then the epidemic quickly sets on, and larger $R_0$'s result in a more rapid propagation. As a rough approximation, we may estimate that it is required that a fraction 
\begin{equation}
    p=1-1/R_0
\end{equation}
of the population is immune to the disease to halt its propagation \citep{Royal2020}. This would give a first guess on when herd immunity will be achieved \citep{Royal2020}, and eventually the fraction of the population that is immune, $i$, may go above this threshold $p$.

The reproduction number informs us about the average number of secondary infections caused by a single infectious individual. However, to give a complete picture of the spread of the disease we must also include time. The generation time, $\tau$, is the average time since a person is infected until he/she infects other people \citep{Nishiura2010}. Then, at the beginning of the outbreak the instantaneous rate of the exponential growth of cases is given by \citep{Royal2020}
\begin{equation}
    r=\frac{R_0-1}{\tau}.
\end{equation}
(This $r$ should not be confused with the fraction of recovered people. We have used the same letter for both concepts trying to keep the same notation as in the standard literature.) Under this model, the number of infectious people would grow as $i(t)\propto \exp(rt)$. However, an exponential growth is not the only possibility and a power law, $i(t)\propto t^\alpha$, has also been proposed \citep{Katul2020}.

Finally, a rather intuitive measure for an epidemic is the doubling time, that is, the time required to double the number of infected cases \citep{Royal2020}
\begin{equation}
    d=\frac{\log{2}}{r}.
\end{equation}

From the point of view of propagation (ecological success of the pathogen), it is much more effective to have a high probability of passing among individuals (high $\beta$, for instance, pathogens that travel through aerosols or respiratory droplets propagate faster than pathogens that require contact with body fluids or fecal-oral transmission) and a low probability of removing the individual from the infectious pool (low $\gamma$). Actually, pathogens that kill their host very quickly (high $\gamma$, like Nipah virus) may not cause large epidemic outbreaks (although as we will see below, this statement has to be modulated by the progression of the disease within the individual, see the digression about the infectivity curve in Sec. \ref{sec:reality}). Typical $R_0$ values for some diseases are: measles (12-18), mumps (10-12), rubella (6-7), Covid19 (3-6), AIDS (2-5), influenza and hepatitis C (1.5-3), Nipah virus (0.5) \citep{Pitlik2020}. 

Whichever $R_0$ is, this model predicts that the infection will vanish when $t$ goes to infinity, $i(\infty)=0$, and that everyone has either been removed (gained immunity or died) or remained susceptible. The proportion of removed and susceptible people in the limit depends on $R_0$, and it is called the final size problem. It can be shown that for the SIR model, the final size, $s(\infty)$, fulfills the equation \citep{Brauer2008}
\begin{equation}
    \log\frac{s(0)}{s(\infty)}=R_0(1-s(\infty))
    \label{eq:finalSize}
\end{equation}

\begin{figure}
    \centering
    \includegraphics[width=8cm]{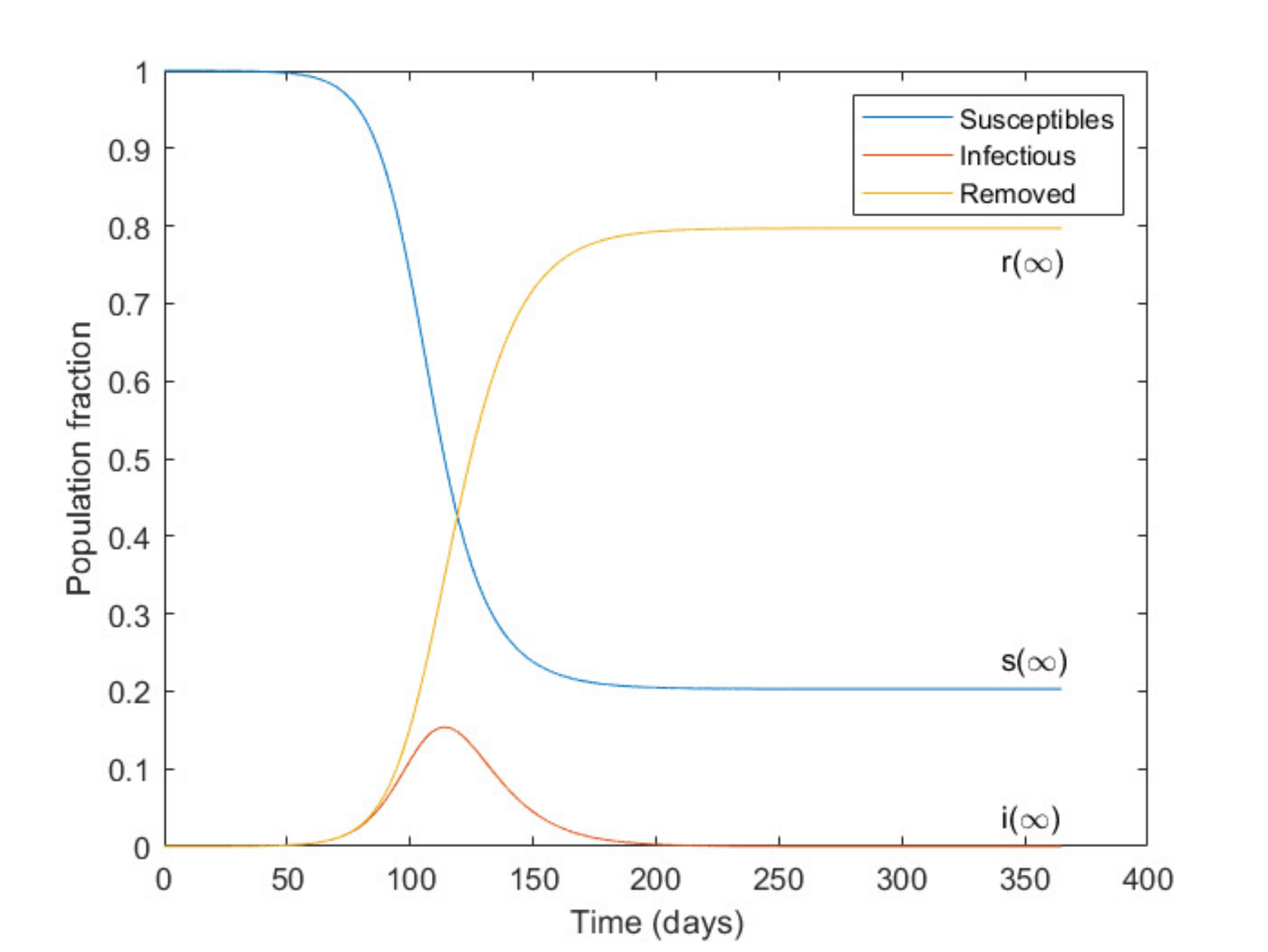}\\
    \includegraphics[width=8cm]{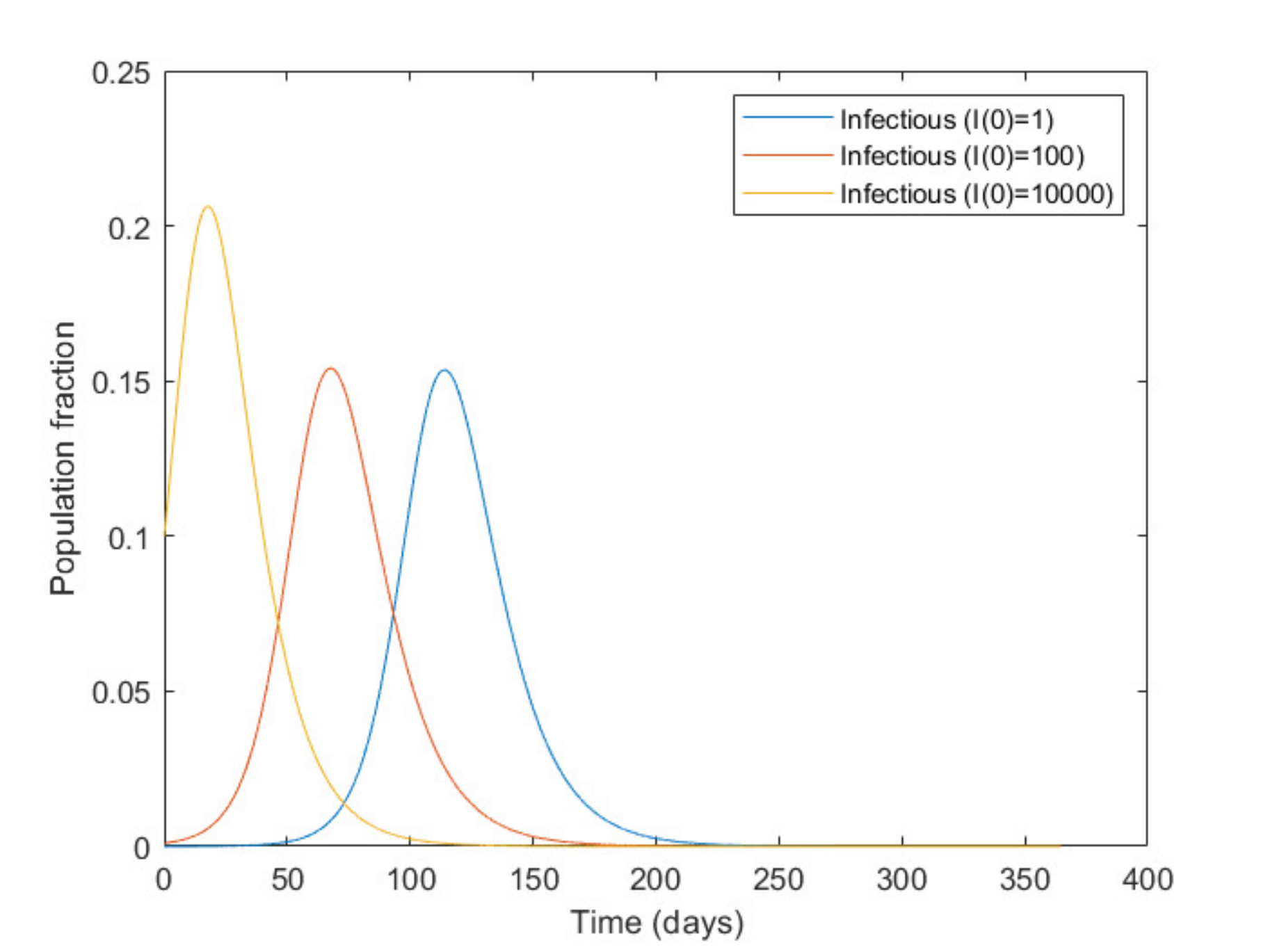}\\
    \includegraphics[width=8cm]{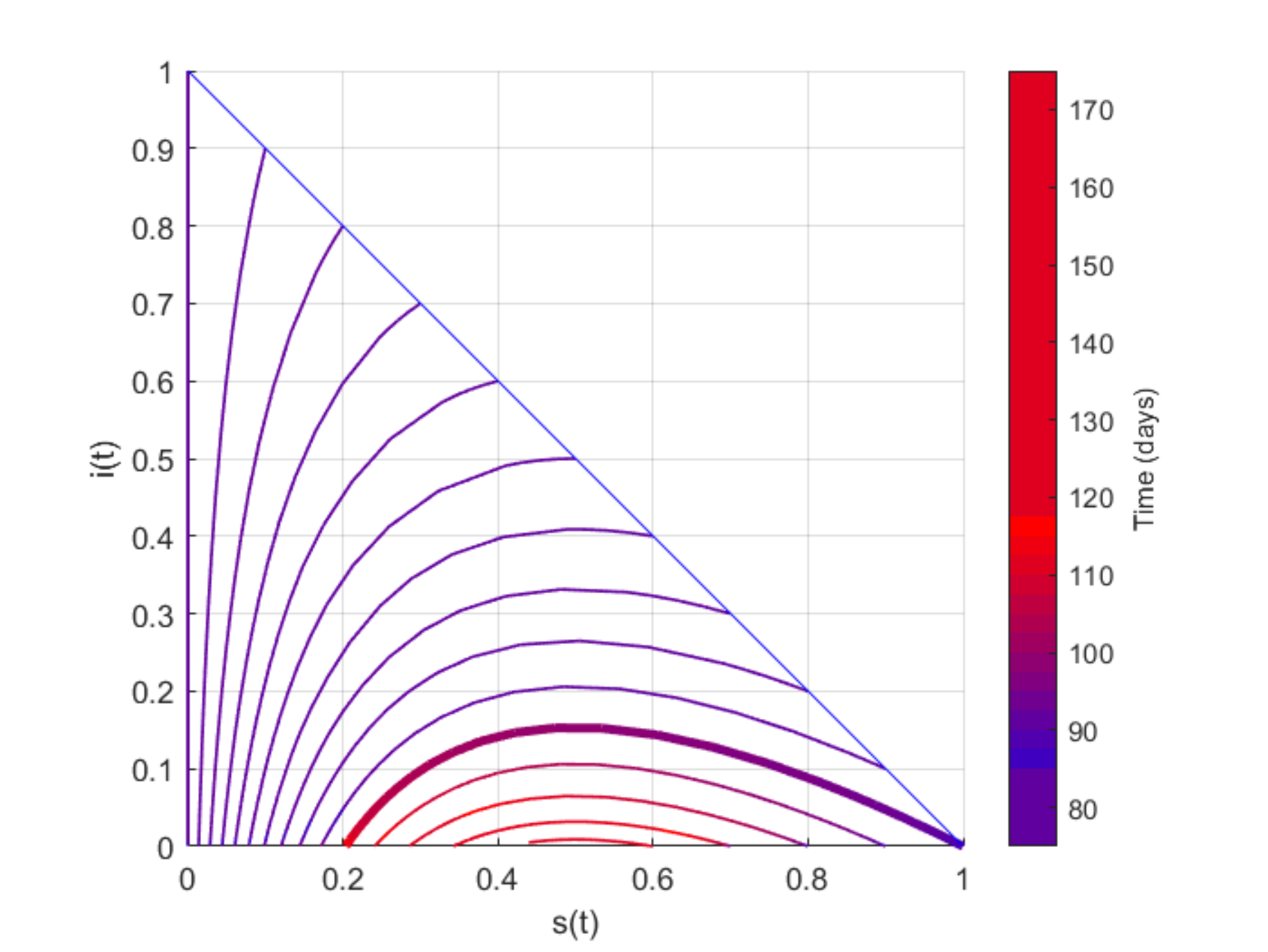}\\
    \caption{Top: Example of SIR process with $\beta=1/5$, $\gamma=1/10$, and just one infectious person at $t=0$ in a population of $N=100,000$ individuals. Middle: Differences in the evolution of infectious individuals as a function of the number of initial infectious ($i(0)$). Bottom: Phase plane of the solutions of the SIR model for $\beta=1/5$ and $\gamma=1/10$.}
    \label{fig:SIR}
\end{figure}

Fig. \ref{fig:SIR} top shows an example of the SIR model's evolution for a particular choice of $\beta$, $\gamma$, and initial conditions. In the figure, it is clear that about 20\% of the population never gets infected, that is $s(\infty)$. Fig. \ref{fig:SIR} middle shows the differences of the evolution depending on the initial number of infectious individuals. It mainly causes an advance of the events and a relatively small effect on the final size (see Eq. \ref{eq:finalSize}). Fig. \ref{fig:SIR} bottom shows the equation system's solutions phase plane. The bold line corresponds to the solution represented at Fig. \ref{fig:SIR} top. Below this bold line, we have the solutions when part of the population is initially immune to the disease (naturally or artificially through vaccination, prophylaxis, or better information leading to better habits).

This model assumes that there are no births, no deaths due to reasons other than the infection being considered, no migration and that the population is homogeneously mixed. The lack of births and deaths for other reasons prevent this model from being used in long-term analyses, in which these considerations are relevant. Another important observation on the SIR model is that, despite its simplicity, it is governed by a set of non-linear differential equations due to the term $is$. This limits very much the mathematical tools available to analyze the system, for instance, tools like the Fourier transform cannot be used.

Despite its limitations, compartmental models are very useful to understand the rationale behind disease propagation (\cite{Brauer2008}) and to evaluate the effect of different public health interventions as was done with SARS (\cite{Gumel2004}), ebola (\cite{Hart2019}), and Covid19 (\cite{Hernandez2021}). 

\section{Deterministic, compartmental models}
\label{sec:compartmental}

\begin{figure*}
    \centering
    \includegraphics[width=0.8\textwidth]{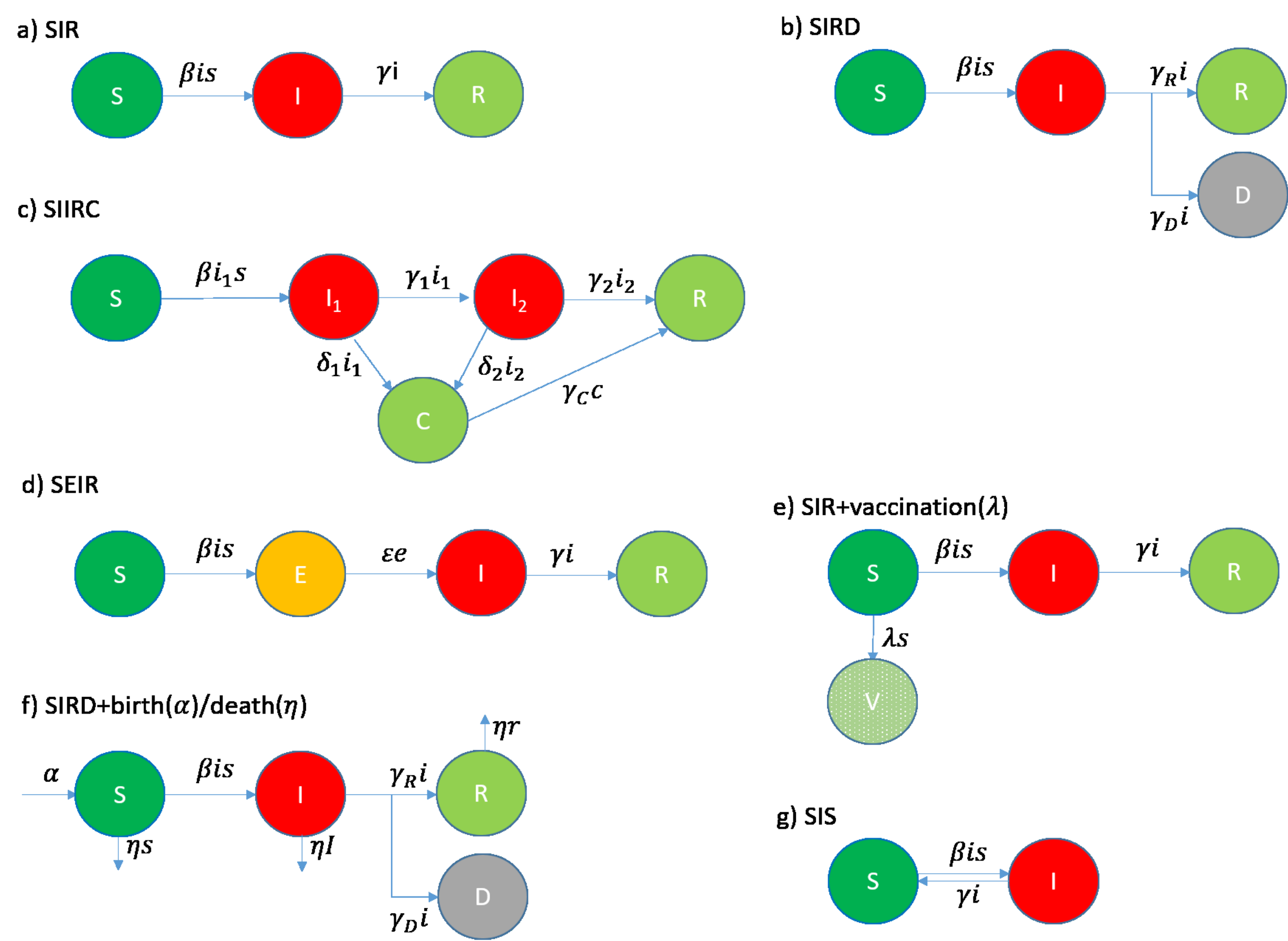}
    \caption{Some deterministic, compartmental models. See text for an explanation.}
    \label{fig:compartmental}
\end{figure*}

The SIR model is probably the best-known deterministic epidemiological model. We may represent it in a system diagram like the one shown in Fig. \ref{fig:compartmental}a. This model is also called compartmental because we have three state variables, also called compartments (susceptible, infectious, removed), and the differential equations govern the dynamics between the three. The transition rate between compartments is shown in the edges between the different compartments.

We may extend this basic model to include more subtle effects:
\begin{itemize}
    \item SIRD: We may distinguish the recovery rate ($\gamma_R$, resulting in the Recovered compartment) from the death rate ($\gamma_D$, resulting in the Deceased compartment), Fig. \ref{fig:compartmental}b.
    \item SIIRC: We may separate the infectious group into two subgroups: one with no or mild symptoms ($I_1$) that can inadvertently propagate the disease, and those with severe symptoms ($I_2$) that are put under control ($C$) \citep{Hart2019}, Fig. \ref{fig:compartmental}c. Actually, this strategy of subdividing in multiple subcompartments can be applied to any one of the compartments \citep{Royal2020}.
    \item SEIR: We may also allow for a latency period in which the person has been infected, but he/she is not infectious. This compartment is called Exposed, Fig. \ref{fig:compartmental}d. The prevalence of a disease in a population is defined as the fraction of active cases, that is, $e+i$.
    \item SIR+vaccination: We may model the effect of vaccination on the system, Fig. \ref{fig:compartmental}e. If we consider all the issues of Sec. \ref{sec:reality}, then we understand the importance of global efforts in vaccination as the only way of preventing a disease from coming back to a region in which it has disappeared (even more crucial for diseases with a high $R_0$ like measles (\cite{Utazi2021})).
    \item SIRD+birth/death: For very long-term epidemiological studies, demographic changes are of interest. In its simplest form, we may include the contribution of a fixed birth rate ($\alpha$) and a fixed death rate ($\eta$, deaths by causes other than the infection being considered), Fig. \ref{fig:compartmental}f. Obviously, the constraint of the population size being constant does not apply as, depending on $\alpha$ and $\eta$, the population may increase or decrease over time.
    \item SIS: We may even consider that after recovery from the disease, the person does not gain immunity or this is not permanent, and the person is susceptible again of going down with the disease, Fig. \ref{fig:compartmental}g. This is the case of common colds (caused by rhinoviruses and coronaviruses) or infections caused by macroparasites (helminths and protozoa).
\end{itemize}
The interested reader is referred to \cite{Hethcote2000} and \cite{Brauer2008} for an extensive review of the mathematical properties of deterministic, compartmental models.

Another extension of the basic SIR model is by taking fractional derivatives in the equation system (\cite{Zhang2020c})

\begin{equation}
    \begin{array}{rcl}
         d^{\alpha_1}s/dt^{\alpha_1}&=&-\beta is \\
         d^{\alpha_2}i/dt^{\alpha_2}&=&\beta is - \gamma i \\
         d^{\alpha_3}r/dt^{\alpha_3}&=&\gamma i \\
    \end{array}
    \label{eq:SIRfractional}
\end{equation}
Alternatively, we could substitute the terms $\beta is$ and $\gamma i$ by any other function (for instance, $\beta i^{\alpha_1} s^{\alpha_2}$, making $\beta$ to depend on time as in $\beta(t)=\beta_0((1-\phi) \exp(-qt)+\phi)$, or any other sensible modification, \cite{Brauer2008}). The goal of all these extensions is to gain model flexibility (new parameters) to better reproduce the observed data.

\begin{figure}
    \centering
    \includegraphics[width=0.65\textwidth]{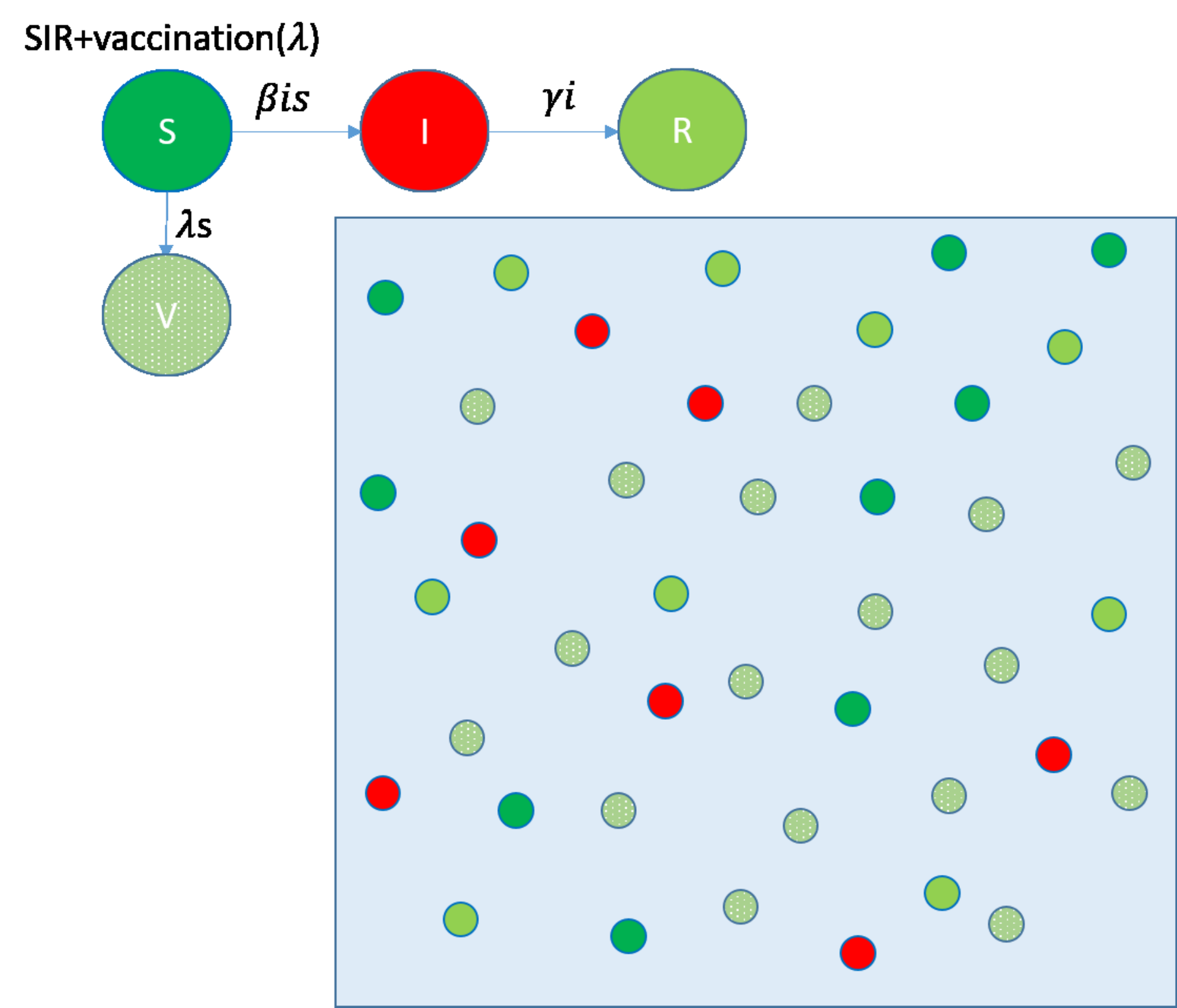} \\    \includegraphics[width=0.65\textwidth]{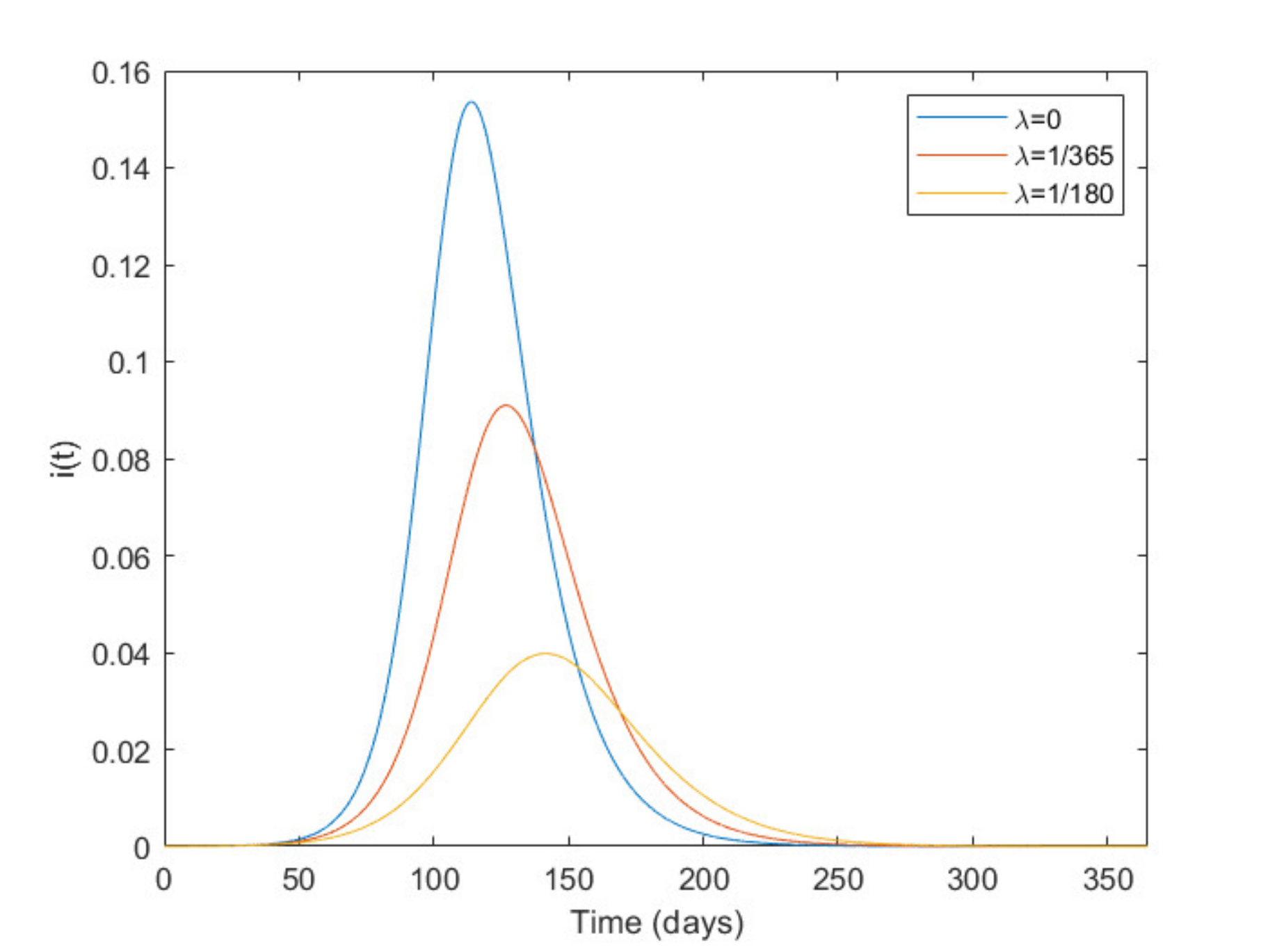}
    \caption{Top: In a population with immunity (either naturally acquired or by vaccination), the probability of encounters between susceptible and infected people decreases. If the effective reproduction number goes below 1, that is, the expected number of cases directly generated by one infected case is less than one, then the infection eventually disappears from the population. Bottom: Example of the evolution of the infected fraction, $i(t)$, if there is no vaccination ($\lambda=0$), and there is vaccination at two different rates ($\lambda=1/365$ and $\lambda=1/180$). In all cases we have used $\beta=1/5$ and $\gamma=1/10$.}
    \label{fig:contagion}
\end{figure}

The basic reproduction number in Eq. \ref{eq:R0} seemed to be a natural definition at the sight of the SIR evolution in Eq. \ref{eq:SIR}. Note that it is dimensionless. It should be understood as the expected number of cases directly generated by one infected case in a population where all individuals are susceptible to infection, that is, there is no immunity by recovery from the disease or by vaccination. However, if we include these other effects, we have an effective reproduction number that depends on time, $R(t)$. The effective reproduction number can be calculated as \citep{Garnett2005}
\begin{equation}
    R(t)=R_0s(t)
\end{equation}
That is, we need to multiply $R_0$ by the fraction of the susceptible population. \citet{Cori2013} provided a method to estimate $R(t)$ directly from data of an ongoing epidemic. When $R(t)<1$ the epidemic starts to shrink. An interesting consequence is that it is unnecessary to vaccinate the whole population to prevent a contagious disease propagating. The reason is that the probability of contact between an infectious and a susceptible person goes down with time (see Fig. \ref{fig:contagion} top). Fig. \ref{fig:contagion} bottom shows the effect of two different vaccination programs (one faster than the other). Regarding vaccination speed, it is essential to make it relatively fast \citep{Forni2021}. If not, the virus may have time to mutate among the active individuals and render the vaccines less effective or, even, ineffective \citep{Kimman2009, Lapinski2012, Tregoning2021}. Another consequence of modelling is that we may adapt our vaccination rate to the proportion of infectious people. If the number of infectious people decreases, we may also lower the vaccination rate without compromising the population safety but reducing the vaccination program cost. We may also reduce the vaccination cost if we address first the risk groups as they have a higher probability of acquiring the disease or if we vaccinate first groups of people with a higher number of contacts. All these criteria are purely epidemiological, but other criteria may also be important, like vaccinating first the most vulnerable groups or the health workers as they are key to keeping the vaccination program. Maybe the most famous case of disease eradication through vaccination was smallpox, which in the 20th century was thought to kill around 300 million people and was officially declared eradicated in 1980. The cost of this vaccination program worldwide was estimated to be only 300M\$ \citep{Barrett2007}.

The calculation of $R_0$ depends on the chosen model (\cite{Brauer2008}). For instance, in the SIRD+birth/death model (Fig. \ref{fig:compartmental}e), it can be shown that the basic reproduction number has to be redefined to
\begin{equation}
    R_0=\frac{\beta}{\gamma}\frac{\alpha}{\eta(1+\eta/\gamma)}
\end{equation}
The disease tends to disappear from the population if $R_0<1$. However, if $R_0>1$, then the disease becomes endemic as long as there is at least one infected case, $i(0)>0$. It can be shown that when time goes to infinity, we have an endemic equilibrium given by \citep{Hethcote2000}
\begin{equation}
    \left(s(\infty),i(\infty),r(\infty)\right)=\left(\frac{\gamma+\eta}{\beta}, \frac{\eta}{\beta}(R_0-1),
    \frac{\gamma}{\beta}(R_0-1)\right)
\end{equation}
Nowadays, several diseases are considered endemic in some world regions, like malaria \citep{Wang2020c}, AIDS \citep{Assefa2020}, or hepatitis B \citep{Shan2018}. Some others, like syphilis or measles, used to be endemic, but effective treatments have successfully brought them down to sporadic outbreaks. \citet{May2001} analyzed the properties of an epidemic from the connectivity properties of a scale-free network. They showed that the disease progression under a SIR model has totally different behavior if the population is very large (infinite) or finite, and that many of the mathematical properties attributed to a SIR model (like the threshold behavior associated to the basic reproduction number, $R_0$) emerge from the scale-free network with a large heterogeneity of the connectivity distribution. This is another indication of the fact that controlling the diseased and susceptible populations is crucial to prevent the contagious agent from spreading.

Despite the importance of $R$ as a way to forecast the evolution of an epidemic, there are several caveats hidden in this number. First, it is an average and, as such, it hides the variability and the shape of the distribution of the number of infected people per infectious person \citep{Lloyd2005}. Significant is the presence of superspreaders (the right tail of the distribution), people who contact many other people due to their lifestyle or job \citep{Wong2020}. Actually, for most infectious diseases with a good track of its spread, it has been verified that most of the infections are caused by a small fraction of the infectious individuals \citep{Royal2020}. Second, it is also a regional average that hides the existence of local infection clusters. Third, it reports the spread of the epidemic but not the severity of the disease (it does not report other important variables like the occupancy of hospital beds, number of deaths, disease sequels, etc.).

We may drop the need for homogeneous mixture of the susceptible and infectious individuals by including the spatial variables in the model, $s(x,y,t)$, $i(x,y,t)$, and $r(x,y,t)$. One of the most common models is the SIRS model that is a reaction-diffusion equation \citep{Gai2020}
\begin{equation}
    \begin{array}{rcl}
         \frac{ds}{dt}&=&-\beta is + \mathrm{div}\left(\mathbf{D}_s \nabla s\right)\\
         \frac{di}{dt}&=&\beta is - \gamma i + \mathrm{div}\left(\mathbf{D}_i \nabla i\right)\\
         \frac{dr}{dt}&=&\gamma i +\mathrm{div}\left(\mathbf{D}_r \nabla r\right)\\
    \end{array}
    \label{eq:SIRS}
\end{equation}
where $\nabla$ is the gradient operator in the spatial coordinates ($x$ and $y$), $\mathrm{div}$ is the divergence operator (also in the spatial coordinates), and $\mathbf{D}_s$ (analogously for $\mathbf{D}_i$ and $\mathbf{D}_r$) is a spatial and time-dependent matrix, $\mathbf{D}_s(x,y,t)$, that locally describes the diffusivity of the different subpopulations. This is an anisotropic, time-variant diffusion. In case that the diffusion matrices do not change over time and space and that the diffusion is isotropic, the matrices can be taken out of the divergence as a constant, and the whole term becomes a Laplacian ($\mathrm{div}\left(\mathbf{D}_s \nabla s\right)=\mathbf{D}_s \nabla^2 s$). This model is very similar to the SIR model of Eq. \ref{eq:SIR}, but we have added the spatial diffusion of the different variables. An interesting consequence of this spatial model is the prediction of the existence of waves that travel through space and time as has been the case of many endemic diseases (maybe, one of the most known waves are the ones of influenza that travel from the North to the South hemispheres and back every year) \citep{Li2009}. \citet{Fontal2021} approached the problem of waves by making $\beta$ to depend on space and time avoiding, in this way, the spatial derivatives.

We may include spatial dependence in a more sophisticated way. Instead of working with population fractions, $(s(t), i(t), r(t))$, let us now work with the absolute number of individuals, $(S(t), I(t), R(t))$, assumed to be continuous variables. We may discretize these variables at spatial locations, denoted by some index $s$. Let us refer to the number of individuals at location $s$ as $N_s(t)$, and the corresponding susceptibles, infectious, and removed individuals as $(S_s(t), I_s(t), R_s(t))$. Let us define a weight matrix $W(s,s')$ that expresses the relationship between the spatial regions $s$ and $s'$ (the rows of the $W$ matrix must add up to 1). Then, we may rewrite the differential equations describing the time evolution of each of the population states at every location $s$ as \citep{Kiss2017}
\begin{equation}
    \begin{array}{rcl}
         \frac{dS_s}{dt}&=&-\beta\left(\sum\limits_{s,s'}W(s,s')\frac{I_s S_{s'
        }}{N_s}\right)\\
         \frac{dI_s}{dt}&=&\beta\left(\sum\limits_{s,s'}W(s,s')\frac{I_s S_{s'
        }}{N_s}\right)-\gamma I_s\\
         \frac{dR_s}{dt}&=&\gamma I_s\\
    \end{array}
    \label{eq:SIRS2}
\end{equation}
This is a large equation system that has to be solved simultaneously for all spatial locations.

\section{System identification}

To fully understand an epidemic situation we must correctly identify the model and its parameters ($\beta$, $\gamma$, $\eta$, $\lambda$, $\mathbf{D}_s$, $\mathbf{D}_i$, $\mathbf{D}_r$, $W(s,s')$, ...). This is a system identification problem. However, as opposed to physical systems in which the sampling process is well characterized in terms of the definition of the measurement, sampling time, and noise statistical properties, epidemiological data quality is much less accurate due to: the lag in the data collection (what is more, this data has to be reported by a distributed network of official agencies, each one with its own variable lag); heterogeneity of the reported data at the level of detail as well as the process being measured (e.g., within the same country, some local governments may report only those cases diagnosed at hospitals, while others may include those cases diagnosed at doctors' offices or other test campaigns; the definition of a simple event as death due to the disease may not be as simple as it appears: it is not the same dying of a disease as dying with a disease, and the technical distinction between the two is not always clear); test samples being biased towards close contacts of infected people; etc. In the best of the cases, these problems result in a large uncertainty of the estimated model parameters, and in the worse case to biased estimates. \citet{Thompson2018} discusses the consequences of choosing the wrong model in the Ebola outbreak between 2013-2016. To maximize the data quality, there has been some debate on the importance of creating reliable national institutions
that homogenize the data allowing proper modeling of the progression of a disease. These surveillance institutions are crucial for the forecast of the progression of diseases \citep{Buckee2020, Cyranoski2021}.

\begin{figure}
    \centering
    \includegraphics[width=0.65\textwidth]{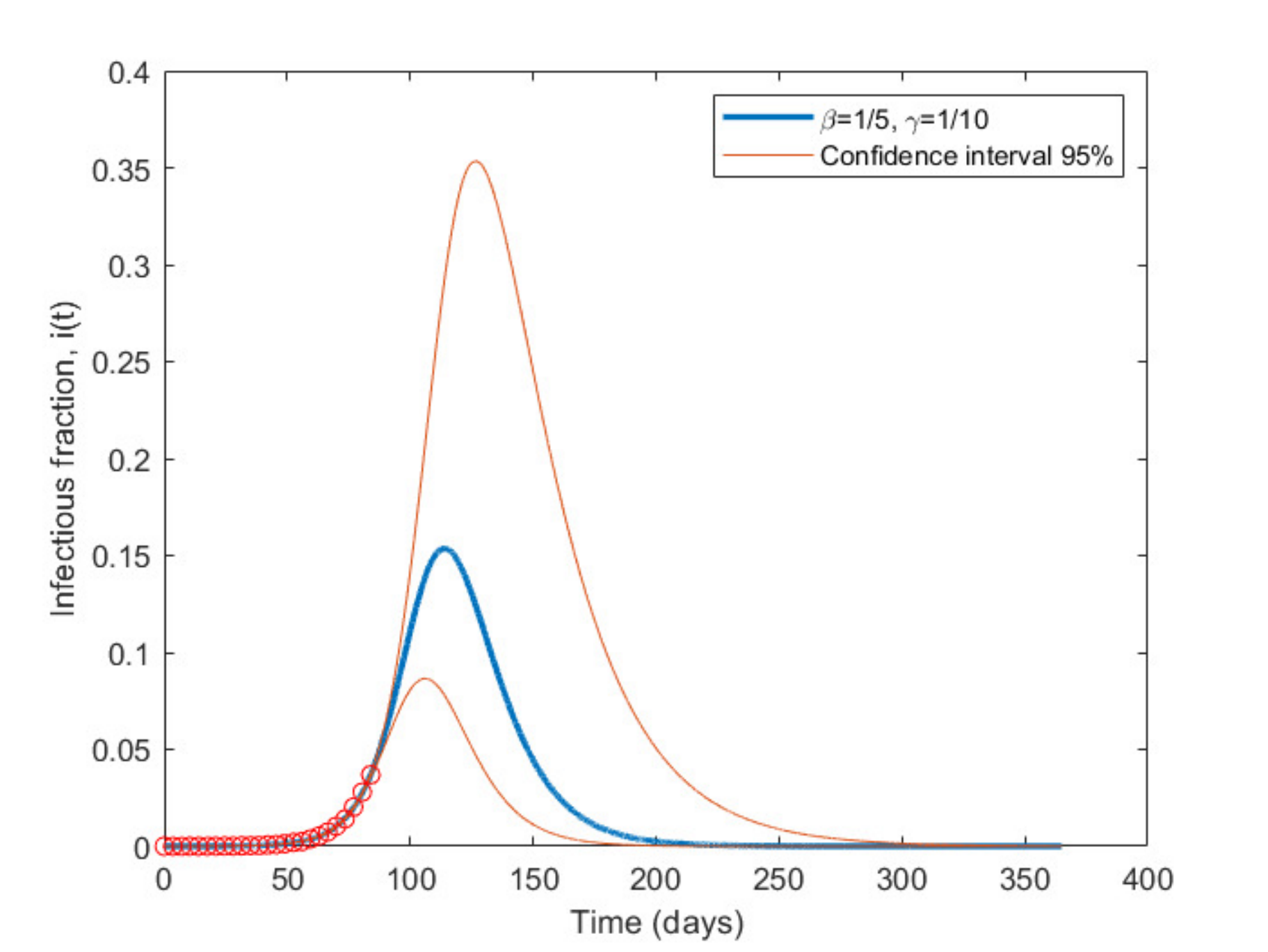}
    \caption{Representation of samples of the fraction of infectious population,
    the actual model ($\beta=1/5, \gamma=1/10$), and the 95\% confidence interval of the prediction.}
    \label{fig:prediction}
\end{figure}

Even in the best case of very accurate data and model, the compartmental models' predictive power is rather limited, especially at the early stages of the epidemic \citep{Castro2020}. The reason is that small errors in the curve fitting translate into large prediction errors due to the multiplicative nature of the underlying equation system. In Fig. \ref{fig:prediction} we show samples from the same model as in Fig. \ref{fig:compartmental}, and the 95\% confidence interval of the predictions. The confidence interval was obtained by bootstrapping the input samples, and the average $R^2$ of the fitting was 0.9996. This example illustrates how cautious we should be with predictions based on the early stages of an epidemic, despite the general public and governments' obvious interest to know its future extent. Additionally, \citet{Odriscoll2021} showed that in these early moments many of the methods available to estimate the reproduction number tend to overestimate its value.

\section{Stochastic processes}
\label{sec:stochastics}

The deterministic, compartmental descriptions of the previous section are rather intuitive and relatively easy to handle mathematically. However, they do not represent contagions' underlying physics: individual people who get infected and removed one by one. Formulating the problem as a continuous fraction of the whole population so that we can use differential equations is a useful trick and relatively accurate when $s$ and $i$ are well separated from the 0 or 1 extremes. When few individuals are infectious at the onset of the disease, continuous models based on differential equations are relatively inaccurate, and stochastic models should be the choice. Actually, the deterministic model is a simplification of the stochastic model. It is assumed that the number of susceptibles, infectious, and removed individuals coincide with the expected values of these random variables divided by the population's total size. Models based on stochastic processes are very useful to model the behavior of the spread early on during an outbreak when the number of infectious individuals is small and there is a possibility of stochastic fade-out.

These models are based on Continuous Time Markov Chains (CTMC; also related to continuous time branching processes in which an infective individual gives raise to several infections (branches)) that are a collection of discrete random variables $\{I(t)\in \mathbb{N}\cup\{0\}: t \in [0,\infty)\}$ that fulfill the Markov property, that is, for any collection of time points $t_0, t_1, ..., t_{n+1}$ such that $0\leq t_0 < t_1 <...< t_{n+1}$, we have
\begin{equation}
    \begin{array}{l}
    \mathrm{Pr}\{I(t_{n+1})=i_{n+1}|I(t_n)=i_n,..., I(t_0)=i_0\}=\\
    \mathrm{Pr}\{I(t_{n+1})=i_{n+1}|I(t_n)=i_n\}
    \end{array}
\end{equation}
$I(t)$ is the number of infectious people at time $t$, and the time points $t_n$ are the instants at which there is a change in the number of infectious people. Note that stochastic processes work with the number of individuals fulfilling a condition instead of the population's fraction fulfilling that condition. An stochastic process is stationary if for all $\tau_1<\tau_2$
\begin{equation}
    \mathrm{Pr}\{I(\tau_2)=j|I(\tau_1)=i\}=p_{ij}(\tau_2-\tau_1)
    \label{eq:pji}
\end{equation}
that is, the probability of going from $i_1$ infectious people at time $\tau_1$ to $i_2$ infectious people at time $\tau_2$ depends only on the time difference, $\Delta \tau$. Note that the $\tau$ times are not restricted to the change time points, $t_n$. Stationary CTMC stochastic processes are characterized by a transition matrix $\mathbf{P}(\Delta \tau)$ such that its $ij$-th element is $p_{ij}(\Delta \tau)$.

One of the most interesting features to study is the time between events. Let us imagine that at time $t_n$ we change to $i$ infectious people. If we make a Taylor expansion of the probability of staying in the same state after a time $\Delta \tau$, we would have
\begin{equation}
    p_{ii}(\Delta \tau)=1-\lambda_i\Delta\tau+o(\Delta \tau)
\end{equation}
where $\lambda_i$ is the coefficient that accompanies $\Delta \tau$ in this Taylor expansion, making it explicit that this coefficient depends on $i$. Many time distributions would be consistent with this Taylor expansion. Among them, one of the simplest is the exponential. In this way, it is often assumed that the distribution of the interevent time when there are $i$ infectious people is an exponential distribution of parameter $\lambda_i$
\begin{equation}
    \Delta \tau \sim \mathrm{Exp}(\lambda_i)
    \label{eq:interevent}
\end{equation}
As a consequence, the average time between events and its standard deviation is $1/\lambda_i$. This process is called a Poisson point process. We may generalize this situation to an arbitrary probability distribution, $f_{\Delta \tau}(\Delta \tau)$. This is exactly the problem addressed by the renewal theory \citep{Smith1958}. It can be proved that the expected number of Infectious cases can be calculated by a convolution called the renewal equation \citep{Champredon2018}
\begin{equation}
    E\{I(t)\}= R_0S(t)\int\limits_0^\infty I(t-\Delta \tau)f_{\Delta \tau}(\Delta \tau)d\Delta \tau
\end{equation}
This equation was first introduced by Euler and is largely used in demographic studies. This connection to the renewal equation has been used in Covid19 to estimate the reproduction number \citep{Pasetto2021}.

To construct a stochastic SIR model, we need to expand the concept of CTMC from one to multiple random processes. A multidimensional CTMC is a collection of discrete random vectors $\{(S(t),I(t))\in (\mathbb{N}\cup\{0\})^2: t \in [0,\infty)\}$ (for simplicity of notation we already particularize the definition to the variables required by a SIR model). We note that, if the population size $N$ is fixed, then $R(t)$ can be automatically computed thanks to the restriction $S(t)+I(t)+R(t)=N$. The transition probabilities have to be redefined to be two-dimensional
\begin{equation}
\begin{small}
    \begin{array}{l}
    \mathrm{Pr}\{S(\tau_2)=s_1+\Delta s, I(\tau_2)=i_1+\Delta i|S(\tau_1)=s_1,I(\tau_1)=i_1\}=\\
    \left\{\begin{array}{rc}
        \frac{\beta}{N}i_1s_1\Delta \tau + o(\Delta \tau) &
           (\Delta s, \Delta i)=(-1,1) \\
        \gamma i_1\Delta \tau +o(\Delta \tau) &  (\Delta s, \Delta i)=(0,1) \\
        1-i_1\left(\frac{\beta}{N}s_1+\gamma\right)\Delta\tau + o(\Delta \tau) &
        (\Delta s, \Delta i)=(0,0) \\
        o(\Delta \tau) & \mathrm{otherwise}
    \end{array}\right.
    \end{array}
    \label{eq:pji2}
\end{small}
\end{equation}
The first line corresponds to an infection, the second line is a removal, the third line reflects the absence of change, and the fourth line implies that there cannot be any other possibility other than the three above (in this model, it is assumed that the time difference is so small that only one event occurs during this short period). At this point, we may give a more general definition of $R_0$ as the average number of infected people by a single infective individual when the whole population is susceptible
\begin{equation}
R_0=\sum\limits_{k=0}^\infty{k}\mathrm{Pr}\{
       S(\infty)=N-k, I(\infty)=k| S(0)=N-1,I(0)=1\}
\end{equation}

\begin{figure}
    \centering
    \includegraphics[width=0.65\textwidth]{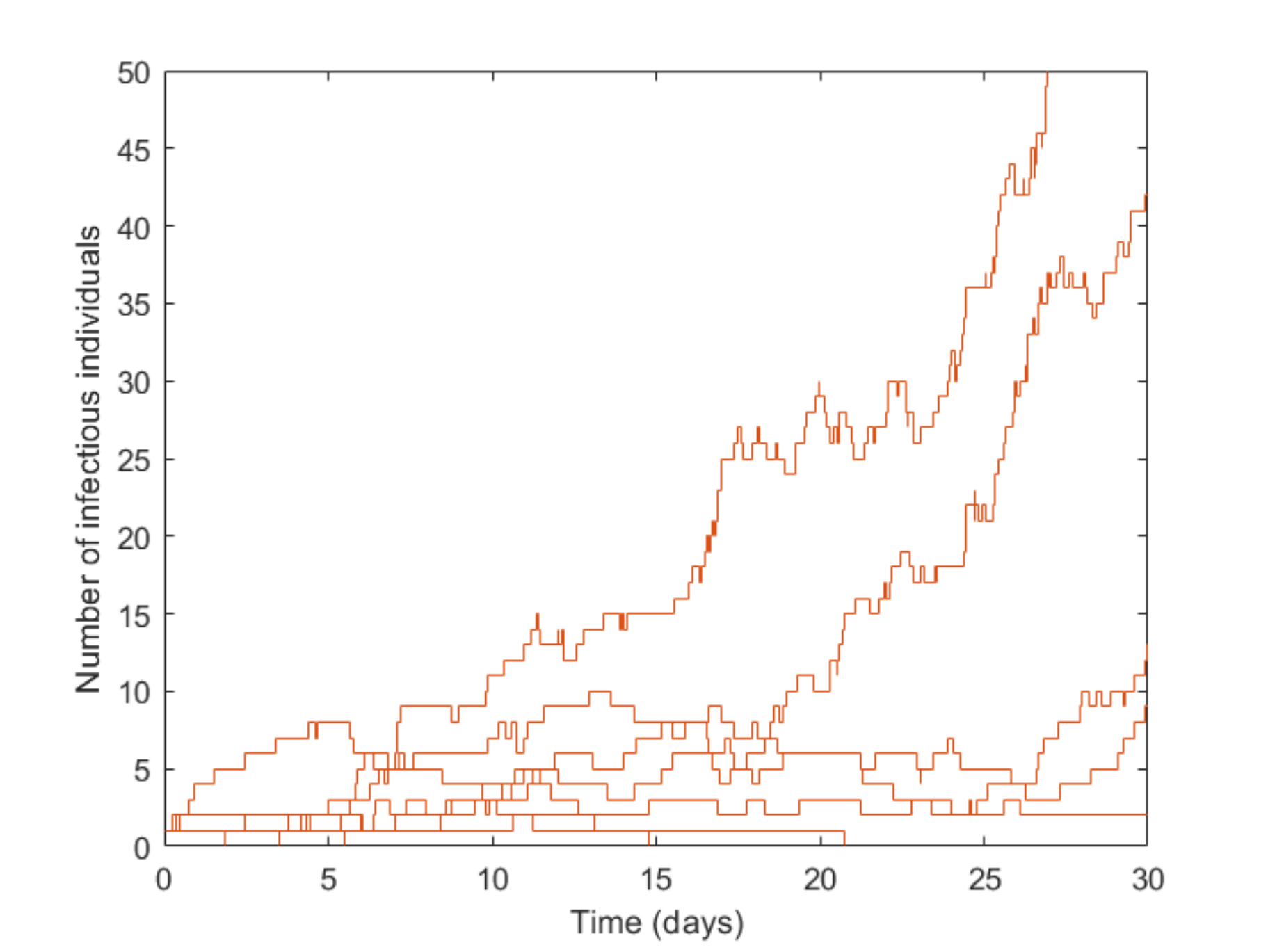}
    \caption{Example of 10 possible realizations of the stochastic process with the same parameters as in Fig. \ref{fig:SIR}. Only the first 30 days are shown. }
    \label{fig:stochastic}
\end{figure}

The stochastic model has an advantage over its differential equation counterpart: at the beginning of an epidemic, when $I(0)=i_0$ is small, there is a probability that the infectious individuals are removed (die or recover) before they have a significant amount of encounters to infect susceptible individuals. In that case, the epidemic ends quickly (this is called a minor epidemic). The probability of this event is 1 if $R_0<1$ and $(1/R_0)^{i_0}$ if $R_0>1$.
Fig. \ref{fig:stochastic} shows some random realizations of the first 30 days of the infection process shown in Fig. \ref{fig:SIR}. Note that in some of the realizations, the epidemic naturally vanishes due to the lucky lack of effective contacts between infectious and susceptible individuals (\cite{Tritch2018}).
For this reason, it is essential to control a disease at its beginning when there are very few infectious individuals. If the number of infectious individuals grows, then there are many more infectious seeds. Due to the equation's multiplicative nature (the term $i_1s_1$ in Eq. \ref{eq:pji2}) the epidemic rapidly grows.

Another interesting remark is that given that there are currently $s_1$ susceptible and $i_1$ infectious individuals, the probability that the next state is a new infection
\begin{equation}
    \frac{\beta/N i_1s_1}{\beta/N i_1s_1+\gamma i_1}=\frac{\beta/N s_1}{\beta/N s_1+\gamma}
\end{equation}
and the probability that it is a recovery
\begin{equation}
    \frac{\gamma i_1}{\beta/N i_1s_1+\gamma i_1}=\frac{\gamma s_1}{\beta/N s_1+\gamma}
\end{equation}

Finally, the interevent time follows an exponential distribution that at time $\tau_1$ has the parameter
\begin{equation}
    \lambda_{\tau_1}=i_1\left(\frac{\beta}{N}s_1+\gamma\right)
    \label{eq:interevent2}
\end{equation}

We may extend these models to account for more complicated real-life effects. For instance, we could model superspreading events by drawing $\beta$ in Eq. \ref{eq:pji2} from a power-law distribution of parameters $\beta_0$ and $\alpha$
\begin{equation}
    \mathrm{Pr}\{\beta > b\} = k \left(\frac{b}{\beta_0}\right)^{-\alpha}
\end{equation}
where $k$ is a constant to make the expression above represent a probability distribution (\cite{Fukui2020}).

Stochastic models were used to study the spread of influenza in the United States and Great Britain \citep{Ferguson2006}. In this work, they made a very thorough analysis of the effect of different public health interventions, and they repeated a similar analysis for the Covid19 pandemic \citep{Ferguson2020}.

\section{Stochastic Differential Equations}

It can be shown that a set of stochastic differential equations can approximate the Markov Chain described in Eq. \ref{eq:pji2} (\cite{Buckingham2018})
\begin{equation}
    d\mathbf{X}(t)=\mathbf{F}(\mathbf{X})dt+\sqrt{\mathbf{V}(\mathbf{X})}d\mathbf{W}
    \label{eq:SDE}
\end{equation}
where
\[\begin{array}{c}
\mathbf{X}(t)=\begin{pmatrix}
   S(t) \\ I(t)\end{pmatrix}, \quad
  \mathbf{F}(\mathbf{X})=\begin{pmatrix}
    -\frac{\beta}{N}IS \\
    \frac{\beta}{N}IS-\gamma I
  \end{pmatrix}, \\
  \mathbf{V}(\mathbf{X}) = \begin{pmatrix}
    \frac{\beta}{N}IS & -\frac{\beta}{N}IS \\
    -\frac{\beta}{N}IS & \frac{\beta}{N}IS+\gamma I
  \end{pmatrix}
  \end{array}
\]
and $\mathbf{W}=(W_1(t),W_2(t))^T$ are two independent Wiener processes that can be thought of as random noise that excites the system. We may compare this equation system with the one of Eq. \ref{eq:SIR}. We see that the latter is simply the deterministic approximation of the set of stochastic differential equations. Formulating the disease spread problem in this new setup allows new understanding of the statistical properties of the stochastic process (stationarity, independence, separability, etc.).

The set of stochastic differential equations above is non-linear due to the term $IS$, and this non-linearity complicates the mathematical extraction of properties from the solution to the equation system. A linear noise approximation can alleviate this (assuming that the $S(t)$ and $I(t)$ are composed of the superposition of a deterministic component, $(\mu_S(t),\mu_I(t))$, and a small stochastic component, and disregarding all stochastic terms with a degree larger than 1) (\cite{Buckingham2018}). The result is that the solution $\mathbf{X}(t)$ of the approximated stochastic differential equation is a multivariate Gaussian process whose mean $\boldsymbol{\mu}=(\mu_S(t),\mu_I(t))^T$ and covariance matrix $\Sigma$ obey the differential equations
\begin{equation}
    \begin{array}{rcl}
        d\boldsymbol{\mu} &=& (\mathbf{A}\boldsymbol{\mu}+\mathbf{b})dt \\
        d\Sigma &=&(\mathbf{A}\Sigma +\Sigma \mathbf{A}^T+\mathbf{U})dt
    \end{array}
\end{equation}
where
\[
    \begin{array}{c}
        \mathbf{A}=\begin{pmatrix} 
           -\frac{\beta}{N}\mu_I(t) & -\frac{\beta}{N}\mu_S(t) \\
           \frac{\beta}{N}\mu_I(t) & \frac{\beta}{N}\mu_I(t)-\gamma \end{pmatrix}, 
        \quad 
        \mathbf{b}=\begin{pmatrix}
           \frac{\beta}{N}\mu_I(t)\mu_S(t) \\
           -\frac{\beta}{N}\mu_I(t)\mu_S(t)
        \end{pmatrix} \\
        \mathbf{U}=\begin{pmatrix} 
           \frac{\beta}{N}\mu_I(t)\mu_S(t) &
           -\frac{\beta}{N}\mu_I(t)\mu_S(t) \\
          -\frac{\beta}{N}\mu_I(t)\mu_S(t) &
          \frac{\beta}{N}\mu_I(t)\mu_S(t)+\gamma \mu_I(t)
    \end{pmatrix}
    \end{array}
\]
The advantage of this simplified formulation is that it is much easier to sample from a Gaussian distribution whose mean and covariance matrix are known than numerically solving the stochastic differential equation in Eq. \ref{eq:SDE}.

\section{Gaussian process regression}

Following the connection above with Gaussian process, we now introduce a different family of models based on Gaussian process regression. They have been used, for instance, to predict the Crimean–Congo hemorrhagic fever (\cite{Ak2018}). These models differ from those described so far in that they do not explain the underlying forces that determine the evolution of the number of susceptibles, infectious, and removed individuals. Instead, they try to explain the number of cases, $y$, as a function of some sensible predictors, $\mathbf{x}\in\mathbb{R}^p$. These predictors may include the time of the measurement (for instance, the month for long studies with periodic behavior), its spatial location, or even weather or environmental measurements. This is a clear advantage over the previous models, which cannot include in the dynamics variables out of the system internal states. The disadvantage of these regression models is that they cannot predict the evolution of a disease in conditions different from the ones in which they have already been observed. An analogy to understand this difference would be trying to model the evolution of a falling object from the forces acting on it ($F=m (d^2 x(t))/(dt^2)$) or from a generic regression polynomial ($x(t)=a_0+a_1t+a_2t^2$). Both models can successfully explain the results of a given experiment. But the first one can additionally be used to predict the results in different experimental settings.

Given a collection of measurements pairs $(\mathbf{x}_i,y_i)$ ($i=1,2,...,N$), let us construct the observation vector $\mathbf{y}=(y_1,y_2,...,y_N)^T$. It is supposed that $\mathbf{y}$ can be explained by a Gaussian process
\begin{equation}
    \mathbf{y}=\mathbf{f}(\mathbf{x})+\boldsymbol{\epsilon}
\end{equation}
where $\boldsymbol{\epsilon}$ are the residuals of the regression (assumed to follow a zero-mean, Gaussian distribution, $\boldsymbol{\epsilon}\sim N(\mathbf{0},\sigma^2_{\epsilon}I$) and $\mathbf{f}$ is a Gaussian process with mean $\boldsymbol{\mu}_f$ and covariance matrix $\mathbf{K}$ given by
\begin{equation}
    \mathbf{K}=\begin{pmatrix}
       k(\mathbf{x}_1,\mathbf{x}_1) & k(\mathbf{x}_1,\mathbf{x}_2) & ... & k(\mathbf{x}_1,\mathbf{x}_N) \\
       k(\mathbf{x}_2,\mathbf{x}_1) & k(\mathbf{x}_2,\mathbf{x}_2) & ... & k(\mathbf{x}_2,\mathbf{x}_N) \\
       ... & ... & ... \\
       k(\mathbf{x}_N,\mathbf{x}_1) & k(\mathbf{x}_N,\mathbf{x}_2) & ... & k(\mathbf{x}_N,\mathbf{x}_N) \\
    \end{pmatrix}
\end{equation}
where $k$ is a kernel function that computes the similarity between two predictor vectors. We will expand later the concept of this kernel.

For predicting the value at a new point $\mathbf{x}_*$, we have that the expected value of $y_*$ is 
\begin{equation}
    \mu_{y_*}=\mathbf{k}_*^T(K+\sigma_{\epsilon}^2I)^{-1}\mathbf{y}
\end{equation}
being
\[\mathbf{k}_*=\begin{pmatrix}k(\mathbf{x}_*,\mathbf{x}_1) & k(\mathbf{x}_*,\mathbf{x}_2) & ... & k(\mathbf{x}_*,\mathbf{x}_N) \end{pmatrix}^T\]

Typical kernels are the Gaussian kernel
\[\begin{array}{rcl}
   k_\Sigma(\mathbf{x}_i,\mathbf{x}_j)&=&(2\pi)^{-p/2}|\Sigma|^{-1/2}\\
   & &\exp\left(-\frac{1}{2}(\mathbf{x}_i-\mathbf{x}_j)^T\Sigma^{-1}(\mathbf{x}_i-\mathbf{x}_j)\right)\end{array}\]
or the Mat\'ern kernel of parameters $\sigma$, $\nu$, and $\rho$
\[k_{\sigma,\nu,\rho}(\mathbf{x}_i,\mathbf{x}_j)=\sigma^2\frac{2^{1-\nu}}{\Gamma(\nu)}\left(\sqrt{2\nu}\frac{d}{\rho}\right)^\nu K_\nu\left(\sqrt{2\nu}\frac{d}{\rho}\right)\]
where $d$ is the distance between $\mathbf{x}_i$ and $\mathbf{x}_j)$, and $K_\nu$ is the Bessel function of the second kind and order $\nu$. If $\nu$ is of the form $\nu=k+1/2$ with $k=0,1,2,...$, then the Mat\'ern kernel can be expressed as the product of a polynomial and an exponential. For instance, for $\nu=5/2$ (a very common choice), we have
\[k_{\sigma,5/2,\rho}(\mathbf{x}_i,\mathbf{x}_j)=\sigma^2\left(1+\sqrt{5}\frac{d}{\rho}+\frac{5}{3}\frac{d^2}{\rho^2}\right)\exp\left(-\sqrt{5}\frac{d}{\rho}\right)\]
We may also include periodicity in the kernel as in
\[k_{\sigma,\omega_0,\rho}(\mathbf{x}_i,\mathbf{x}_j)=
\sigma^2\exp\left(-\frac{\sin^2(\omega_0 d)}{\rho}\right)\]
This is useful to represent seasonal outbreaks.

We may decompose the kernel into several pieces depending on the nature of the predictors. For instance, if the predictors include time, spatial, and environmental variables ($\mathbf{x}=(\mathbf{x}_{time},\mathbf{x}_{space},\mathbf{x}_{env})$), then we may measure the similarity of its different components using different kernels
\begin{equation}
    \begin{array}{rcl}
    k(\mathbf{x}_i,\mathbf{x}_j)&=&k_{time}(\mathbf{x}_{time,i},\mathbf{x}_{time,j})+\\
    & &k_{space}(\mathbf{x}_{space,i},\mathbf{x}_{space,j})+\\
    & &k_{env}(\mathbf{x}_{env,i},\mathbf{x}_{env,j})
    \end{array}
\end{equation}

In this kind of models, the regression coefficients are the parameters that define the similarity kernel (e.g., $\Sigma$, $\sigma$, or $\rho$; the researcher typically fixes $ \nu$). These parameters are estimated by Maximum Likelihood on the current set of observations ($\{(\mathbf{x}_i,y_i)\}$).


\section{Bayesian regression models}

The last family of methods we will review assumes \textit{a priori} distributions for the different elements that describe the events taking place during the disease's spread. These distributions are normally fixed, and their parameters must be determined either from experimentally observed data, or fitted by some kind of optimization procedure (more details are given below). This new family would be a generalization of the Stochastic models presented in Sec. \ref{sec:stochastics}.

As an example, we will show a simplified model of the one presented in \cite{Flaxman2020} for Covid19. Let us assume that we can only observe the number of deaths, $d_n$ from the disease on the $n$-th day ($n=0,1,2,...$). It is assumed that this number of deaths follow a negative binomial distribution with a mean $\mu^d_n$ (that we will model later) and variance $\mu^d_n+(\mu^d_n)^2/\eta_1$, where $\eta_1$ is a random variable distributed as a half-normal $N^+(0,\theta_1)$ (if $X$ is normally distributed with mean $\mu$ and variance $\sigma^2$, $X\sim N(\mu,\sigma^2)$, then $Y=|X|$ is said to follow a half-normal distribution of parameters $\mu$ and $\sigma^2$, $Y\sim N^+(\mu,\sigma^2)$).

Let us now consider the time from infection of a particular person to his/her death, $t^d$. This variable is decomposed in two periods: from the infection to the onset of the symptoms, and from the onset of the symptoms to death. Each of these times is supposed to follow a $\Gamma$ distribution with parameters $(\theta_2,\theta_3)$ and $(\theta_4,\theta_5)$, respectively. Consequently $t^d$ follows a distribution that is the convolution of the two $\Gamma$'s: $t^d\sim \Gamma(\theta_2,\theta_3)\star \Gamma(\theta_4,\theta_5)$. Given this distribution of the time from infection to death, we could compute the probability of dying on the $n$-th day after infection as $\pi^d_n=\mathrm{Pr}\{n\leq t^d <n+1\}$

The next element to model is the infection-fatality-rate, $\gamma^d$, that is the probability of death given infection. This $\gamma^d$ is supposed to follow a log-normal distribution of parameters $\theta_6$ and $\theta_7$.

The mean of the negative binomial is modeled as a discrete convolution between the sequences $i_n$ (the number of new infections on the day $n$) and $\pi^d_n$
\begin{equation}
    \mu^d_n=\gamma^d \sum\limits_{n'=0}^{n-1} i_{n'} \pi^d_{n-n'}
    \label{eq:meanDeaths}
\end{equation}
This convolution is simply the sum of the new infections on the days before $n$ multiplied by the probability that those infections die on the day $n$.

Finally, we must model the number of new infections on the $n$-th day. This is done by modeling the generation time, $t^g$. This is the time between a person gets infected and the moment at which he/she passes the infection to another susceptible individual. This time is supposed to follow another $\Gamma$ distribution of parameters $\theta_8$ and $\theta_9$. The probability that a person communicates the disease exactly $n$ days after being infected is $\pi^g_n=\mathrm{Pr}\{n\leq t^g<n+1\}$. Then, the number of new infections is
\begin{equation}
    i_n=\left(1-\frac{\sum\limits_{n'=0}^{n-1}i_{n'}}{N}\right)R_n \sum\limits_{n'=0}^{n-1} i_{n'} \pi^g_{n-n'}
\end{equation}
The first term in parenthesis reflects the depletion of the susceptible pool of individuals (the total size of the population is $N$). The second term, $R_n$, is a time-varying reproduction number that may include the effect of interventions like school and university closures, self-isolation if ill, forbidding public events, lock-downs, etc. Finally, the sum has a similar interpretation as in Eq. \ref{eq:meanDeaths}, that is, the probability of the infected people from days before $n$ producing a new infection on the day $n$.

The time-varying reproduction number is modeled as
\begin{equation}
    R_n=R_0\exp\left(-\sum\limits_k \alpha_k I_{k,n}\right)
\end{equation}
That is, a basic reproduction number in the absence of any intervention, $R_0$, times a term that reduces this value as a function of the measures adopted. Each measure has an index $k$, reduces by a factor $\exp(-\alpha_k)$, and the variable $I_{k,n}$ is an indicator variable that takes the value 1 if the $k$-th measure has been taken on the day $n$, and 0 if it has not. $R_0$ is a random variable that follows a half-normal distribution of parameters $\theta_{10}$ and $\eta_2$, and $\eta_2$ is another random variable normally distributed with zero mean and variance $\theta_{11}$.

An extension of these models that take into account multiple time series is the one presented in \cite{Ssentongo2021}. This work tries to simultaneously model the number of infections in all African countries. The main two novelties with respect to the models introduced so far are its multivariate and autoregressive nature. The evolution of the number of infections is formulated in the same framework as time-series forecasting problems. In particular, let us call $I_{s,n}$ to the number of infections at country $s$ at time $n$. This number is supposed to be distributed as a negative binomial with mean $\mu_{s,n}$ and overdispersion parameter $\psi$. The mean of the negative binomial depends on the number of previously observed infections and the relationship of that country with the rest of the countries in the time series
\begin{equation}
    \mu_{s,n}=m_{s,n}+\lambda_{s,n}\sum\limits_{d=1}^Du_dI_{s,n-d}+\phi_{s,n}\sum\limits_{d=1}^D\sum\limits_{s'\neq s}u_dW(s,s')I_{s',t-d}
\end{equation}
The first term, $m_{s,n}$, is called the endemic component and it is specific to each country. It can be a constant or modelled in more complex ways to account for the country population, living conditions, daily temperature and humidity, testing policy, government stringency, mobility restrictions, etc. For instance, we could state
\begin{equation}
    \log(m_{s,n})=\alpha_s+\beta_s\log(N_s)+\gamma_sT_{s,n}
    \label{eq:msn}
\end{equation}
where $N_s$ is the population of the country $s$, $T_{s,n}$ is the average temperature of that country on day $n$, and $\alpha_s$, $\beta_s$, and $\gamma_s$ are random effects drawn from Gaussian distributions: $\alpha_s\sim N(\theta_0,\theta_1)$, $\beta_s\sim N(\theta_2,\theta_3)$, $\gamma_s\sim N(\theta_4,\theta_5)$.

The second term, $\lambda_{s,n}...$, gives the autoregressive behavior of a time series on itself. $d$ is a time lag whose maximum, $D$, should cover the whole period from the appearance of symptoms in an individual and the appearance of symptoms in a secondary infection. The third term, $\phi_{s,n}...$, accounts for the mutual influence of the different spatial locations on each other. The coefficients $\lambda_{s,n}$ and $\phi_{s,n}$ could be simple coefficients to be estimated or follow regression models as the one in Eq. \ref{eq:msn}. Finally, the set of parameters $u_d$ couples the local ($\lambda_{s,n}$) and global ($\phi_{s,n}$) autoregressions.

The model parameters ($\theta_1$, $\theta_2$, ...) are either taken from previous knowledge (scientific literature or experiments specially designed to estimate them) or are simultaneously estimated from the data available. This latter fitting is performed with specialized software (for instance, STAN (\cite{Carpenter2017})) that implements advanced Monte Carlo sampling methods to find the parameters that better represent (Maximum \textit{a posteriori}) the observed data.

As we can see, these models are rather flexible. However, they have several drawbacks: 1) the distribution priors ($\Gamma$, log-normal, etc.) and construction of these priors ($\eta_1, \eta_2$) bring useful constraints if they really represent reality; otherwise, they bias the results; 2) if too many parameters are sought ($\theta_1, ..., \theta_{11}$) the model has too much freedom to explain past data, but it may not predict well the future; 3) if many of these parameters are obtained from the literature, they bring useful information if they really represent reality; otherwise, they bias the results again.

\section{Limitations of the models}
{\label{sec:reality}}

\begin{figure}
    \centering
    \includegraphics[width=0.65\textwidth]{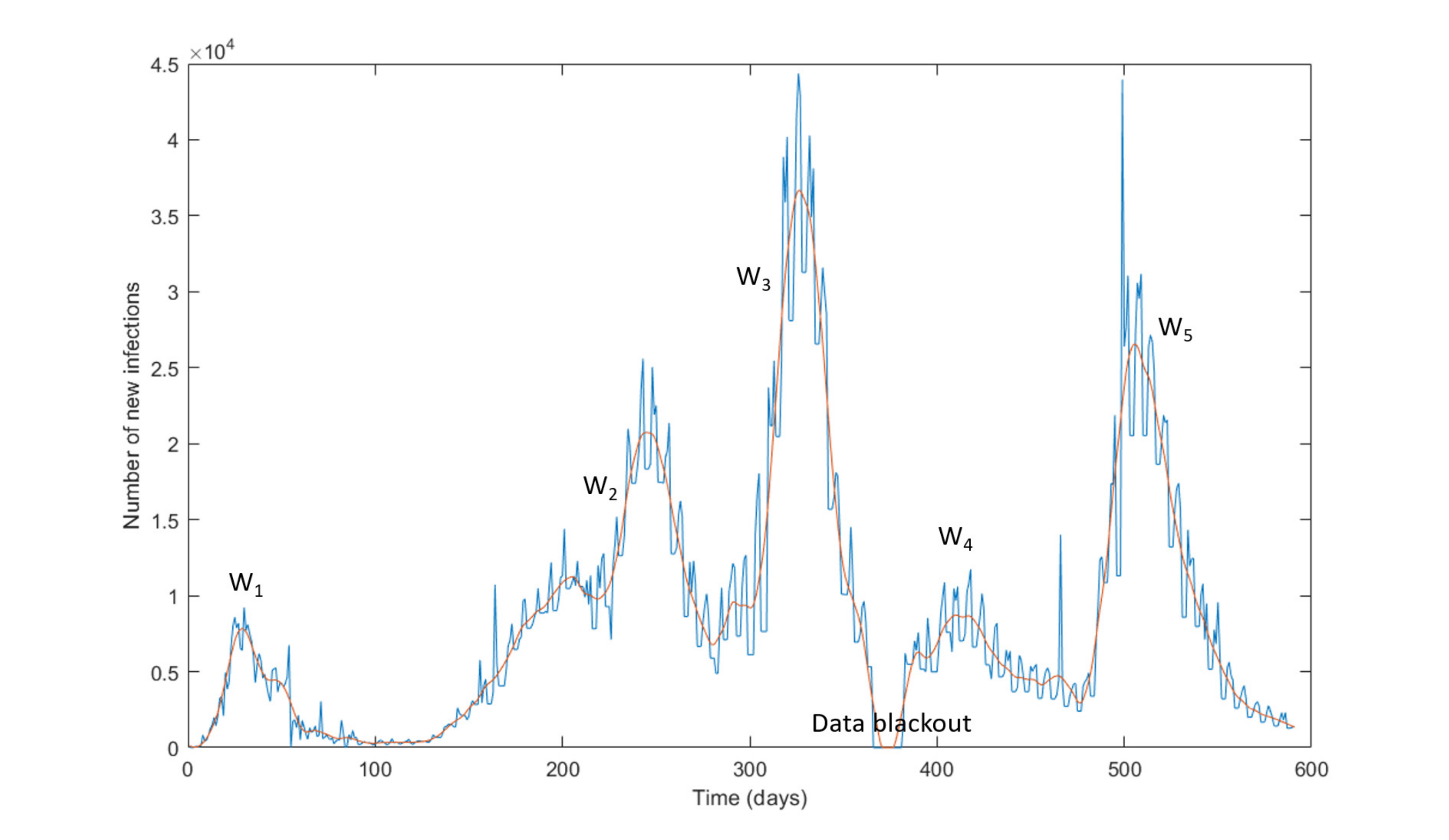}
    \caption{Number of new infections in Spain daily from the day 0 (first reported case). The $W_i$ labels show different infection waves.}
    \label{fig:Spain}
\end{figure}

Deterministic, compartmental models and stochastic processes are useful tools to understand some of the basic properties of the spread of contagious diseases (evolution under controlled conditions, final size, probability of starting an epidemic, existence of waves, etc.). However, to be mathematically tractable, they need to be necessarily simplistic. They do not consider many effects that occur in real life that make the models invalid. For instance, the analysis of the new cases reported daily from the onset of the infection (day 0) in Spain clearly shows multiple waves that cannot be explained by a single SIR model (see Fig. \ref{fig:Spain}). Regression models have the additional limitation that they are valid as long as the conditions under which they were estimated hold valid. If these conditions change, then they do not represent reality any longer.

There can be many reasons for this mismatch between models and reality, but we can summarize them all like the fact that $\beta$ and $\gamma$ change with space, time, or host variables like age or social status (\citet{Verity2020} performed a very detailed analysis of these distributions for Covid19). Any particular choice is just a sample from a statistical distribution (that also changes with space and time) of real-life events:
\begin{enumerate}
    \item Some of the reasons are obvious, like public health interventions that include: case detection, border controls, area quarantine, blanket travel restrictions, antiviral treatment and prophylaxis, case isolation, household quarantine, reactive school/workplace closure, promoting remote work, regulating public spaces' capacity, social distancing, and vaccination \citep{Ferguson2006}. For instance, \citet{Jarvis2021} studied the effect of social distancing on the reproduction number. Although each one of these interventions can be modelled and their effects evaluated under fixed conditions, in reality the application of these interventions change over space, time, intensity, and adherence making it very difficult to foresee their real effect.
    
    \item The definition of successful infective contact goes well beyond a homogeneous mixture of susceptibles and infectives, and it depends on many factors like: 
    \begin{itemize}
    \item the number of contacts of a particular individual: some people have many more contacts than other people for lifestyle or work reasons, these are the famous superspreaders (\citet{Katul2020} studied the effect of superspreaders on the basic reproduction number for Covid19). On the other extreme, it has been suggested to promote the number of interactions with immune people (serologically tested) as a way to reduce the number of effective infective contacts \citep{Weitz2020}. The time between infections has also its own distribution \citep{Nishiura2010}, which in its turn depends on many other factors as age, social status, etc.
    
    \item the contact patterns of the different individuals (the race, gender, and age of the contacts as these biological characteristics may affect the susceptibility (as is the case of poliomyelitis or rubella, that affect mostly children or young adults, or SARS-CoV2, that children seem to be less susceptible); urban and rural proportions of the population, population density, city structures, and commuting habits as they define the daily flows of individuals (\cite{Bouffanais2020}); societal, school, office, and household structure, as these cultural variables may affect the definition of individual clusters with higher levels of contact between them and lower levels of contacts outside their groups; etc.) \citep{Leung2017}. For Covid19, it has been recognized the increased number of infections occurred in health centers \citep{Royal2020}.
    
    \item pathogen flows due to people migrations, tourism, and work trips (we live in a world that is increasingly interconnected with the number of domestic and international passengers around 4 billion per year (\cite{Becken2020}), and flight routes forming a scale-free network in which diseases can easily travel long distances (\cite{Lau2020})). However, the pathogen flow may be much more complex like the West Nile fever whose path could go from man to mosquitoes, eaten by migratory birds, which travel long distances, or other species acting as natural reservoirs (these cross infections between animals and humans and vice versa are called zoonoses, like swine and birds being the natural reservoir of influenza viruses that may mutate and jump to human hosts; for instance, one of the measures to prevent the propagation of MERS in humans was the vaccination of dromedaries). \cite{Vila2021} studied the parallelism between pathogen spreads and biological invasion by some species into a new ecosystem. Certainly, these two domains (epidemiology and ecology) could have very fruitful cross-fertilization not only because their mathematical tools are similar, but also because animals serve as vectors for the transmission and spread of many human diseases.
    
    \item duration and intensity of the contact (for the transmission of some diseases, it suffices to inhale the air exhaled by someone else, while some others may require a much more intimate contact), the definition of contact may be even more complicated if we take into account that some diseases require intermediate vectors to transmit the pathogen (e.g., bubonic plague requires rat fleas, and malaria needs female anopheles mosquitoes) and that we need then to include ecosystem (or even, perhaps, weather) considerations (\cite{Demers2020}).
    
    \item the term $\beta is$ in the SIR model assumes that an infectious person has more probability of infecting another individual as the number of susceptible individuals grows. Still, this assumption may not be valid, especially for diseases that require close, like sexual, contact (the number of susceptible individuals that a single infectious person can infect does not linearly grow with the number of susceptibles). To avoid this latter problem, some models use the term
    $\beta N^{v-1} is$ where $v$ is a parameter to be estimated, and that may take a value between 0 and 1 (typical values are below 0.1, while the standard SIR formulation corresponds to $v=1$) (\cite{Hethcote2000}).
    \end{itemize}
    
    \item The pathogen is normally not of a single type. All humans belong to a single species, \textit{Homo sapiens sapiens}. Still, we are all different from each other as individuals, and our abilities (our genes and our phenotypes, which genes are active at each time, are different). Similarly, pathogens (viruses, bacteria, protozoa, fungi, algae, lichens, helminths, parasites, etc.) are all different as individuals, most importantly for this article's scope, in their ability to propagate and the severity of the caused disease. These differences are more pronounced in pathogens whose life cycle is shorter (for instance, some viruses are replicated in vesicles called viral factories where thousands of copies of the same virus are performed by a single cell (\cite{Kieser2020}); each of these copies has a small probability of committing copy errors (mutations)). This mutation mechanism is at the basis of the success of life exploring different solutions \citep{Rodpothong2012}. We have witnessed this in SARS-CoV2 \citep{Lythgoe2021}, with different mutations propagating more quickly than others and rapidly monopolizing the infected population (\cite{Trucchi2021, Volz2021}). To grasp the problem's size, we may consider that the viral load (number of copies of the virus) for some respiratory diseases has been reported to be as high as $10^7$ copies/mL in nasopharyngeal fluids \citep{Yoon2020}. If we multiply this number by the total of fluid affected within a person and the number of infected people, we can easily imagine that the number of variants of the virus can rapidly grow. 
    In the case of SARS-CoV2, these changes have been carefully tracked over time (\url{https://www.gisaid.org/phylodynamics/global/nextstrain}), resulting in more than 4,000 major variants in less than a year. This variability implies that the constants $\beta$ and $\gamma$ are no longer constants but a distribution of values depending on the specific pathogens infecting an individual.
    
    \item The same occurs with the infected person. The variability of the immune system among individuals imposes an important source of variation that makes that $\beta$ and $\gamma$ are no longer constants. We have seen that pathogens introduce genetic variations to explore possible more successful individuals in terms of reproduction. Similarly, our immune system has also at its disposal a hypervariable tool that tries to identify foreign infections: antibodies and T-cell activation and expansion. Also, when a disease is largely known in a population, many of its individuals develop immunity, but the same diseases can devastate whole populations if these diseases are new within these groups (this was the case of measles, smallpox, pertussis, and influenza that were brought to America in the XVIth Century by Europeans, and this has been the case of the new Covid19). Differences among individuals in their immune systems' efficiency result in differences in the effectivity transmission and recovery. Moreover, the pathogen load within a person also changes over time. Just after being infected, the person probably does not transmit the disease because the number of copies of the pathogen is not yet high enough. This dependence over time is modeled through an infectivity function (see Fig. \ref{fig:infectivity}) that is different for each individual. Reality can be even more complicated because the same pathogen may stay in the person in an inactive form and only appear at irregular periods (for example, the Herpes Simplex Virus is one of these). In this case, the infectivity function would have as many peaks as active intervals. A very perilous stage is the one between the time in which the person can infect other individuals, but he/she does not show any symptom so that he/she does not take any precaution to avoid contagion. These individuals are called carriers, and they may inadvertently produce a large portion of the infections. This is the case of AIDS in which an infected person takes many years to develop symptoms. Finally, the duration of the immunity within the person is also another important variable as it returns the person to the susceptible group. There are diseases for which immunity lasts forever, while for some other diseases immunity lasts just a few months or years \citep{Amanna2007}.
    
    \begin{figure}
        \centering
        \includegraphics[width=0.65\textwidth]{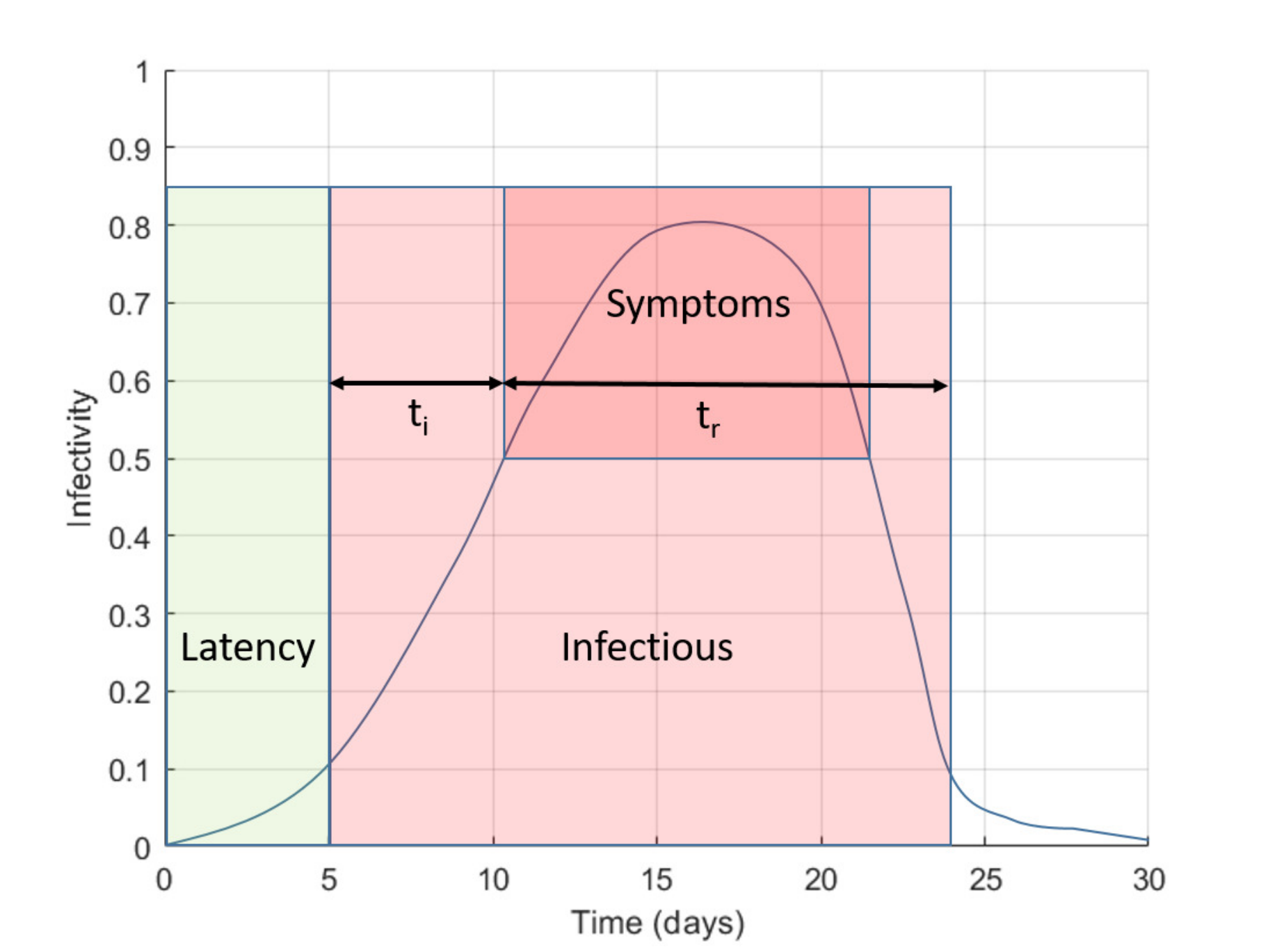}
        \caption{Example of an infectivity function of a single individual. Below a given pathogen load, the person cannot effectively transmit the disease (this time is called latency). Symptoms appear when the pathogen load is sufficiently high, and there might be a period, $t_i$, in which the person is a transmitter of the disease, but he/she does not know because he/she does not have any symptom.}
        \label{fig:infectivity}
\end{figure}

\end{enumerate}

In general, mathematical models of the spread of disease are very useful to simulate the behavior of an epidemic if a given intervention is adopted or if there are specific changes in the pathogen or the host, assuming all other variables are kept fixed. They are also useful to understand the past and tracking an epidemic in real-time. However, their predictive power can be rather questionable due to the space-time-biological-societal changing nature of the problem that cannot be forecasted by the model. This was well illustrated in the case of the Ebola outbreak between 2013-2016. The forecasted number of cases ranged from 6,000 to 10,000 \citep{Thompson2018}, but the true number of cases was about 30,000 \citep{Coltart2017}. This is not a remark on the quality of the work above, but on the difficulty of forecasting disease outbreaks and a warning to the credibility of the forecast attempts at the middle of a pandemic (a search of ``covid and forecast'' in Scopus returns 1,136 journal papers as of Oct. 2021).

\section{Agent-Based Models}

Deterministic or stochastic simple models have the advantage of being mathematically tractable and allowing the derivation of important insights into the dynamics of contagious diseases. However, as we have seen in the previous section, the reality is much more complicated than the simple situations that can be easily modeled. Agent-Based Models (ABMs) aim to fill this complexity gap by simulating agents' behavior over time. An agent can be a simple individual, a family, a school, an institution, or any other sensible entity. An agent has many attributes like health state (in our example Susceptible, Exposed, Infectious, or Removed, but many more possible subdivisions are possible), spatial position, age, economic status, daily schedule (school or work hours, commuting, recreation, and shopping, etc.), household configuration, a social network (friends, colleagues, family, occasional contacts, etc.), main activity (student, job, unemployed, retired, etc.), and any other feature that we find relevant for the study of the disease's propagation. Each feature may have its own dynamics (for short simulations: commuting time or transportation means may change from one day to the next or over weekends, contact duration and intensity may also change, ...; for long simulations: age grows with time, the economic status or social network may vary with age, ...). Even the nature and frequency of the contact can be different (contact duration and intensity at school or work are different from those at public transportation or shops). Local density of people in a given region, land use (business or residential neighborhoods), and daily flow patterns can also be incorporated into the simulation. We should use real data to define all these variables. This data may come from the population census, research studies, shopping patterns, public transport data, mobile phone networks, or any other source. The simulation is performed in time steps. At every step, every agent acts in each one of its features. We may incorporate randomness into the model through random variables for moving from one state to another in any of its characteristics.

\begin{figure*}
    \centering
    \includegraphics[width=0.6\textwidth]{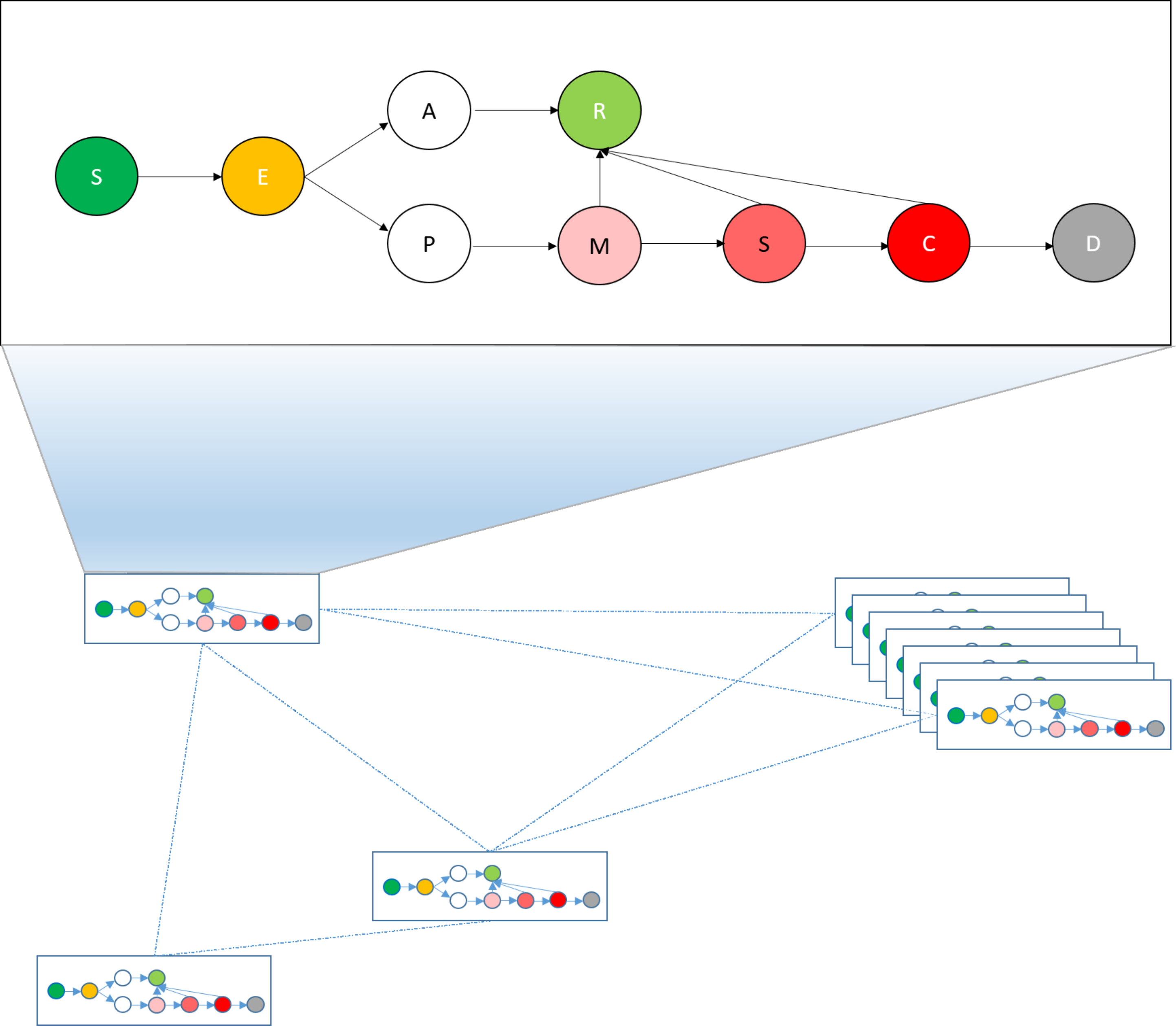}
    \caption{A population of individuals relate to each other (dashed blue lines). Each individual has its own internal state (Susceptible (S), Exposed (E), Asymptomatic (A), Presymptomatic (P), with Mild (M), Severe (S), or Critical (C) symptoms, Dead (D)).}
    \label{fig:ABM}
\end{figure*}

This kind of simulation has been used to study AIDS propagation (\cite{Huang2015b}), measles (\cite{Hunter2018}), or Covid19 \citep{Hernandez2021}. It has even been used to evaluate the effect of different social distancing interventions (\cite{Silva2020}). \citet{Kerr2001} introduced an ABM simulator of Covid19 spread whose parameters can be tuned to the observations of a particular region and, then, different interventions can be simulated to understand their effect. In their simulator, an  agent followed a SEIR path with multiple infectious states (see Fig. \ref{fig:ABM}): first, it was Susceptible, then it becomes Exposed, Asymptomatic or Presymptomatic (with Mild, Severe, or Critical symptoms). Finally, the person, recovers or dies. The transition from one state to the next is governed by a random time variable (in \citet{Kerr2021}, log-normally distributed).

There is no doubt that these micromodeling approaches are the most flexible tools to simulate the evolution of a contagious disease. However, they have two main drawbacks: 1) we need faithful data for each of the agents' attributes; 2) we cannot model large systems and long times due to their high computational cost. The first drawback is rather severe as it is very difficult to obtain the parameters required to represent reality. Interestingly, \cite{Hernandez2021} showed that the SEIR model (see Sec. \ref{sec:compartmental}) can be as accurate as ABMs if the distribution of the incubation and removal times are allowed to be arbitrary instead of the traditionally assumed exponential distribution. A more extensive comparison of ABMs and compartmental models was performed in \citet{Panovska2021}.

\section{Integro-differential models}

Despite the flexibility of ABMs, these models cannot efficiently simulate millions of individuals. For this reason, we must find a compromise between the enormous flexibility of ABMs and the efficiency of compartmental models by dealing with aggregated measures over groups.

We now present a possible extension of the SIR model that gives great flexibility to incorporate many subpopulation and time distribution effects (\cite{Hernandez2021}).
We start from the SEIR model in which individuals spend some time in which they are infected, but they are not infectious yet (see Fig. \ref{fig:compartmental}). The variation of susceptibles is given by
\begin{equation}
    \begin{array}{rcl}
         dS(t)&=&-\beta I(t)S(t)dt \\
    \end{array}
    \label{eq:Sevol}
\end{equation}
Let us assume that they spend a time $t_i$ in the $E$ compartment before becoming effectively infectious (see Fig. \ref{fig:infectivity}). Then, the variation of $E$ is the new susceptibles that have been exposed and infected minus the ones that appeared $t_i$ time units ago because they move to the infectious group.
\begin{equation}
    \begin{array}{rcl}
         dE(t)&=&\beta I(t)S(t)dt - \beta I(t-t_i)S(t-t_i)dt\\
         &=&dS(t)-dS(t-t_i)u(t-t_i)
    \end{array}
\end{equation}
where $u(t)$ is Heaviside's step function. We may extend this delay idea to the other two compartments, considering that the recovery time is fixed, $t_r$:
\begin{equation}
    \begin{array}{rcl}
         dI(t)&=&dS(t-t_i)u(t-t_i)-dS(t-t_r)u(t-t_r)\\
         dR(t)&=&dS(t-t_r)u(t-t_r)\\
    \end{array}
\end{equation}
Now, let us relax the condition that the $t_i$ and $t_r$ times are fixed. We do so by dividing the population into $K$ subpopulations (indexed by $k$) that have their own fixed times (this subdivision is only needed for the $E$, $I$, and $R$ populations)
\begin{equation}
    \begin{array}{rcl}
       dE^{(k)}(t)&=&dS(t)
        -dS(t-t_i^{(k)})u(t-t_i^{(k)}) \\
       dI^{(k)}(t)&=&dS(t-t_i^{(k)})u(t-t_i^{(k)})\\
        & &-dS(t-t_i^{(k)}-t_r^{(k)})u(t-t_i^{(k)}-t_r^{(k)})\\
       dR^{(k)}(t)&=&dS(t-t_i^{(k)}-t_r^{(k)})u(t-t_i^{(k)}-t_r^{(k)})\\
    \end{array}
\end{equation}
We now make the subpopulations infinitely small, to the point in which $t_i^{(k)}$ and $t_r^{(k)}$ become continuous variables, that we will call again $t_i$ and $t_r$. The relative abundance of the different subpopulations can be captured by a probability density function of $t_i$ and $t_r$ ($p_i(t_i)$ and $p_r(t_r)$). Finally, we add (integrate) all the subgroups in a single group ($E$, $I$, and $R$) to get the integro-differential equations
\begin{equation}
    \begin{array}{rcl}
         \frac{dS}{dt}(t)&=&-\beta I(t) S(t) \\
         \frac{dE}{dt}(t)&=&S'(t)-\int_0^t S'(t-t_i)p_i(t_i)dt_i \\
         \frac{dI}{dt}(t)&=&\int_0^t S'(t-t_i)p_i(t_i)dt_i \\
            & &-\int_0^t\int_0^{t-t_i} S'(t-t_i-t_r)p_i(t_i)p_r(t_r)dt_rdt_i
    \end{array}
\end{equation}
where $S'$ is the derivative of $S(t)$ with respect to $t$. It can be proved that the standard SEIR model is obtained when the $t_i$ and $t_r$ distributions are exponential, which was one of the consequences of the simple stochastic model (see Eq. \ref{eq:interevent}) but that does not very often explain the interevent times observed in real epidemics. In \cite{Hernandez2021} it is shown that this general model can mimic the behavior of ABMs with different network properties but at a much lower computational cost.

Finally, we can also relax the condition that all infection rates are fixed, that is, $\beta$ is constant. We do so by considering that the infection rate of a person of type $k'$ infects a person of type $k$ is $\beta_{k',k}$. Then we would rewrite the variation of the susceptibles of type $k$ as
\begin{equation}
    \begin{array}{rcl}
         dS^{(k)}(t)&=&-\sum\limits_{k'}\beta_{k',k} I^{(k')}(t) S^{(k)}(t)dt \\
    \end{array}
\end{equation}
If we assume that the infectivity only depends on the nature of the infectious individual and not the characteristics of the susceptible, then $\beta$ solely depends on $k'$
\begin{equation}
    \begin{array}{rcl}
         dS^{(k)}(t)&=&-\left(\sum\limits_{k'}\beta_{k'}p_{k'}\right)I(t) S^{(k)}(t)dt \\
    \end{array}
\end{equation}
where we have defined $p_{k'}$ as the proportion of individuals in the $k'$-th infectious compartment. If we now consider infinitely small compartments, then $\beta_{k'}$ becomes a random variable with distribution $p_\beta$, and the infection rate in Eq. \ref{eq:Sevol} is just its average value.

\section{Conclusion}

The spread of contagious diseases has been studied from many different perspectives. As opposed to physical systems, disease spread is more difficult to model accurately because, besides the large biological variability of the pathogen and host (which result in a distribution of the underlying parameters, rather than constants), we need to add the variability due to social behavior, regulations, economical and healthcare systems, and the lack of homogeneous and high-quality data. Moreover, these societal variables change very quickly over time. For all these reasons, the identification of the parameters required for the mathematical models is rather complicated. In any case, the predictions based on these models, no matter their mathematical sophistication and their explanatory power to the past, should be taken with care for several reasons: 1) they are normally based on average values of the underlying parameters; 2) these parameters are estimated in very noisy environments, with observed data that vary in quality, definition, and delay, and they consequently have large estimation errors and biases; 3) tractable models are relatively simple, and they do not account for many real-life effects; 4) even, in the best case, the uncertainty of the predictions due to the uncertainty of the parameters can be quite considerable, especially at the early stages of the epidemic. All these problems do not suggest giving up any attempt to foresee the near future and prepare for it as has been shown in multiple attempts addressing Covid19 in the special issue of the Philosophical Trans. Royal Soc. B \citep{Pickett2021}. However, they call for extra caution when dealing with contagious diseases, especially when they are still new within the population.

Probably, the most important lessons to learn from these models is that epidemics are better controlled when: 1) the pathogen is correctly isolated, and its transmission means clearly identified, 2) infectious individuals are quickly spotted (as soon as possible, specially at the onset of the epidemic breakout), and 3) their numbers of potentially infective contacts are minimized (ideally set to 0).

\selectlanguage{english}
\FloatBarrier

\bibliographystyle{apalike}
\bibliography{bibliografia}


\end{document}